\documentclass[12pt]{article}
\usepackage{epsfig,a4wide}

\usepackage{graphicx}
\usepackage[all,2cell,ps]{xy}
\newcommand{\dmu}{\partial_\mu}

\newcommand{\dmuu}{\partial^\mu}

\newcommand{\lc}{{\cal L}}

\newcommand{\oh}{{1\over2}}
\newcommand{\trh}{{3\over2}}

\newcommand{\sqd}{\sqrt{2}}
\newcommand{\sqt}{\sqrt{3}}
\newcommand{\sqs}{\sqrt{6}}
\newcommand{\sqdt}{\sqrt{2 \over 3}}

\newcommand{\squt}{{1 \over \sqt}}
\newcommand{\squd}{{1 \over \sqd}}
\newcommand{\squs}{{1 \over \sqs}}

\title{Dynamically Generated Open and Hidden Charm Meson Systems}

\author{D.~Gamermann$^1$, E.~Oset$^1$, D.~Strottman$^1$ and M. J. Vicente Vacas$^1$ \\ 
{\small{\it $^1$Departamento de F\'{\i}sica Te\'orica and IFIC,
Centro Mixto Universidad de Valencia-CSIC,}}\\
{\small{\it Institutos de
Investigaci\'on de Paterna, Aptdo. 22085, 46071 Valencia, Spain}}
}

\begin{document}

\maketitle

\abstract{We will study open and hidden charm scalar meson resonances within two different models. The first one is a direct application of a chiral Lagrangian already used to study flavor symmetry breaking in Skyrme models. In another approach to the problem a $SU(4)$ symmetric Lagrangian is built and the symmetry is broken down to $SU(3)$ by identifying currents where heavy mesons are exchanged and suppressing those. Unitarization in couple channels leads to dynamical generation of resonances in both models, in particular a new hidden charm resonance with mass 3.7 GeV is predicted. The small differences between these models and with previous works will be discussed.}

\section{Introduction}

 The discovery and soon after confirmation of charmed scalar resonances by BaBar and Belle [\ref{barbar}], [\ref{belle1}], [\ref{belle2}] has opened a controversy about their structures. In the $q\bar q$ picture these resonances are naturally assigned as $^3P_0$ states in the $^{2S+1}l_j$ notation, but calculations done long before in the framework of quark model potentials [\ref{isgur}] had mass predictions which turned out to be more than 100 MeV off the real mass of the states. Lattice calculations also fail in calculating the masses with a $q\bar q$ assignment [\ref{lat1}].

 This situation has sparked the discussion whether these resonances could have a different structure. Some authors have suggested a $qq \bar q \bar q$ structure [\ref{qqqq1}], [\ref{qqqq2}] or a mixing between four quarks and the usual $q \bar q$ structure [\ref{qqqq3}]. Also molecular states have been suggested [\ref{mol1}], [\ref{mol2}], [\ref{mol3}]. For discussions on these and other controversial heavy mesons refer to [\ref{disc1}], [\ref{disc2}], [\ref{disc3}].

 Unitarized coupled channel models have also been considered for the study of these resonances in [\ref{lutz1}], [\ref{lutz2}] and [\ref{chiang1}]. These works have used a chiral Lagrangian based on heavy quark symmetry [\ref{refe1}], [\ref{refe2}], [\ref{refe3}], [\ref{refe4}] for the open charm sector which neglects exchanges of heavy vector mesons in the implicit Weinberg-Tomozawa term. We intend to extend the study for all possible sectors of the interaction, including the hidden charm and the double charmed sector. The exchange of heavy vector mesons is also taken into account in our approach, with the corresponding terms properly accounting for the larger mass of the heavy vector mesons.

 In the present work we will construct a Lagrangian for the interaction of the 15-plet of pseudoscalar mesons in $SU(4)$. $SU(4)$ symmetry breaking will be considered by suppressing the exchange of heavy vector mesons. The $SU(3)$ structure of the interaction will be thoroughly analysed and unitarization in coupled channels will lead to the generation of scalar resonances corresponding to poles in the T-matrix. For comparison, in the open charm sector, we will also solve the problem with yet another Lagrangian which has been considered in the study of flavor symmetry breaking effects in Skyrme models. This Lagrangian gives similar results, supporting our present model.

 The work is organised as follows: in the next section a brief review of the structure of the $SU(4)$ 15-plet will be presented. Section 3 is dedicated to the explanation of the construction of the Lagrangian and in section 4 the theoretical framework for solving the scattering equations in a unitarized approach is presented. Section 5 is dedicated to analysing the results and a brief summary is presented in section 6.

\section{The 15-plet}

 In this work the framework already used to study the interaction of the octet of
 pseudoscalar mesons in $SU(3)$ [\ref{meme1}] will be extended to include charmed mesons. This will involve some extrapolation to $SU(4)$. In the $q \bar q$ picture, mesons involving charm will be classified as $4 \otimes \bar 4=15 \oplus 1$, hence belonging to a 15-plet or a singlet. It is
 interesting to understand how the 15 representation of $SU(4)$ breaks down into representations of
 $SU(3)$ and in which channels the interaction of the multiplets will be attractive or
 repulsive.
 
\begin{figure}
\begin{center}
\includegraphics[scale=0.5,angle=00]{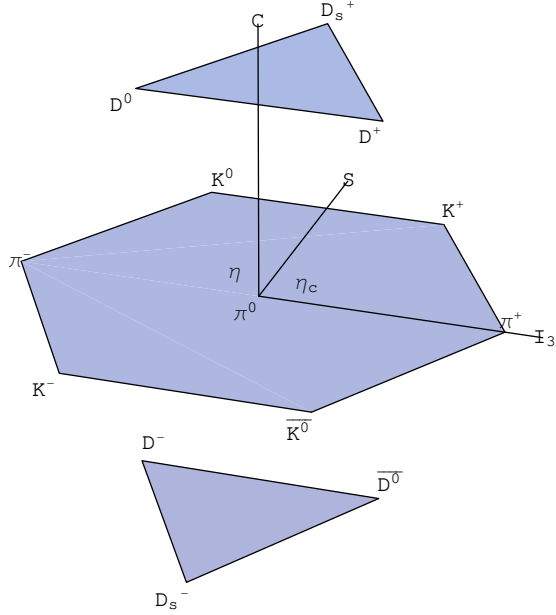} 
\caption{15-plet from $SU(4)$ with its mesons assignments.} \label{15plet}
\end{center}
\end{figure}

 The pseudoscalar mesons are represented by a 15-plet of $SU(4)$ as shown in figure \ref{15plet}. Once $SU(4)$
 symmetry is broken into $SU(3)$, the 15-plet breaks down into four multiplets of the
 lower symmetry, an octet, a triplet, an antitriplet and a singlet:
 
\begin{equation}
15 \longrightarrow 1 \oplus 3 \oplus \bar 3 \oplus 8 .
\end{equation}

 The octet and the singlet have null charm quantum number, the triplet and the antitriplet have
 negative and positive charm quantum number, respectively.
 When studying the meson-meson interaction, one can decompose the scattering of two 15-plets of $SU(4)$ according to its $SU(3)$ inner structure, see Table \ref{decompo}.
 
 \begin{table}
 \begin{center}
 \begin{tabular}{c|l}
 \hline
 charm & Interacting multiplets \\
 \hline 
 \hline
 & \\
 2 & $\bar 3 \otimes \bar 3 \rightarrow 3 \oplus \bar 6$ \\
 \hline
 & \\
 1 & $\bar 3 \otimes 8 \rightarrow \bar 15 \oplus \bar 3 \oplus 6 $ \\
   & $ \bar 3 \otimes 1 \rightarrow \bar 3$ \\
 \hline
 & \\
 0 & $\bar 3 \otimes 3 \rightarrow 8 \oplus 1$ \\
   & $1 \otimes 1 \rightarrow 1$ \\
   & $8 \otimes 1 \rightarrow 8$ \\
   & $8 \otimes 8 \rightarrow 1 \oplus 8_s \oplus 8_a \oplus 10 \oplus \bar {10} \oplus 27$ \\
 \hline
 \end{tabular}
 \caption{$SU(3)$ decomposition of the meson-meson interaction in $SU(4)$. The sectors not shown in the table correspond to the $C=-1,-2$ states which are just charge
 conjugate states (antiparticles) from the ones shown.} \label{decompo}
 \end{center}
 \end{table}

 The interaction $8\otimes 8$ is already very well studied [\ref{meme1}]-[\ref{refee2}]. It is already known that
 in s-wave it generates dynamically poles in the T-matrix which are associated with the $f_0(980)$, $\kappa$,
 $a_0(980)$ and $\sigma$ resonances. Also the interaction $\bar 3 \otimes 8$ has already been studied
 [\ref{lutz1}], [\ref{lutz2}], [\ref{chiang1}] in some different approaches. In this sector, as we will show in the
 following sections, the
 interaction, when diagonalized in a $SU(3)$ basis, is attractive in $\bar 3$ and $6$,
 while repulsive in the $\bar {15}$, so one could expect to generate 5 poles in ($S,I$): from the $\bar 3$
 with isospin $0$ and $1\over 2$ and strangeness $1$ and $0$, respectively; from the $6$ with 
 $I=0,{1 \over 2}, 1$ and $S=-1,0,1$, respectively. Moreover
 the interaction in the $\bar 3 \otimes 3$ is attractive in both the $8$ and the $1$, so one can, in principle,
 expect four new resonances in the hidden-charm sector. The interaction in the $C=2$ sector is repulsive in the
 $\bar 6$ and in the $3$ the interaction vanishes.

 Apart from studying the different sectors separately, it is interesting to see how the mixing
 of states from different sectors with the same $SU(3)$ representation of Table \ref{decompo} affects the interaction.

 Furthermore, if heavy resonances are generated from the $3\otimes\bar3$ one can expect, in
 principle, that the mixing of those heavy channels with light ones coming from $8\otimes8$ will make its width
 quite large because of the large phase-space available for decay. However, we shall also see that there are subtleties in the interaction which suppress these decays.

\section{The Lagrangians}

 The $SU(3)$ lowest order chiral Lagrangian reads [\ref{gasser}], [\ref{chilag}]:

\begin{equation}
\lc_\chi={f_\pi^2\over4}Tr\left(\dmu U\dmuu U\right)+{f_\pi^2 m_\pi^2 \over 4}Tr\left( U+U^\dagger-2\right) \label{chil}
\end{equation}
where $U$ is the field containing the pseudoscalar mesons from the $SU(3)$ octet and $Tr$ represents a trace in flavor space.:

\begin{eqnarray}
U&=&e^{{i\sqrt{2}\phi_8\over f_\pi}} \\
\phi_8&=&\left( \begin{array}{ccc}
 {\pi^0 \over \sqd}+{\eta \over \sqs} & \pi^+ & K^+  \\
 \pi^- & {-\pi^0 \over \sqd}+{\eta \over \sqs} & K^0 \\
 K^- & \bar K^0 & {-2\eta \over \sqs} \\
  \end{array} \right) \label{phi8}
\end{eqnarray}

 Flavor symmetry breaking effects can be introduced with two new terms in the Lagrangian [\ref{flasybre1}], [\ref{flasybre2}]:

\begin{eqnarray}
\lc_{SB}&=&{f_K^2 m_K^2-f_\pi^2 m_\pi^2 \over 6} Tr\left( (\hat 1-\sqrt{3} \lambda_8)(U+U^\dagger-2)\right)+ \nonumber\\
&-&{f_K^2-f_\pi^2 \over 12}Tr\left( (\hat 1-\sqrt{3} \lambda_8)(Ul_\mu l^\mu+l_\mu l^\mu U^\dagger)\right) \label{chisybr}\\
l_\mu&=&U^\dagger \dmu U
\end{eqnarray}
where $\lambda_8$ is the $SU(3)$ generator.

 In [\ref{walliser}] these Lagrangians are extended to $SU(n)$. In this new approach the symmetry breaking sector is written as:

\begin{eqnarray}
\lc_{SB}&=&{1\over8}\sum_{k=3}^n \gamma_k Tr\left((\hat 1-\sqrt{{1\over2}k(k-1)}\lambda_{k^2-1})(Ul_\mu l^\mu+l_\mu l^\mu U^\dagger)\right)\\
&+&{1\over8}\sum_{k=3}^n \delta_k Tr\left((\hat 1-\sqrt{{1\over2}k(k-1)}\lambda_{k^2-1})(U+U^\dagger-2)\right)
\end{eqnarray}
but now $U$ belongs to a $SU(n)$ representation.

By expanding the $U$ matrix until fourth order in the meson fields, one can identify the mass and kinetic terms for each field and fix the symmetry breaking parameters for $SU(4)$ and $SU(3)$ as:

\begin{eqnarray}
\gamma_3&=&{4\over6}(f_K^2-f_\pi^2)\\
\delta_3&=&{4\over3}(f_K^2 m_K^2-f_\pi^2 m_\pi^2)\\
\gamma_4&=&{1\over2}(f_D^2+f_K^2-2f_\pi^2)\\
\delta_4&=&f_D^2 m_D^2-{1\over3}f_K^2 m_K^2-{1\over3}f_\pi^2 m_\pi^2
\end{eqnarray}

 In this work we will consider only the difference between $f_D$ and $f_\pi$ which is about 70\% and we will make the aproximation $f_K=f_\pi$.

For constructing our model we will first consider a $SU(4)$ field containing all fields from the 15-plet\footnote{What here is called $\eta$ and $\eta_c$ are actually $\eta_8$ and $\eta_{15}$ states, which
 mix with a singlet, $\eta_1$ to form the physical states $\eta$, $\eta'$ and $\eta_c$, but in this work this mixing
 won't be taken into account, and it will be considered that the physical states are just described by their most important
 components.}:

\begin{eqnarray}
 \Phi&=&\sum_{i=1}^{15}{\varphi_i \over \sqd}\lambda_i \nonumber =\\ &=&\left( \begin{array}{cccc}
 {\pi^0 \over \sqd}+{\eta \over \sqs}+{\eta_c \over \sqrt{12}} & \pi^+ & K^+ & \bar D^0 \\ & & & \\
 \pi^- & {-\pi^0 \over \sqd}+{\eta \over \sqs}+{\eta_c \over \sqrt{12}} & K^0 & D^- \\& & & \\
 K^- & \bar K^0 & {-2\eta \over \sqs}+{\eta_c \over \sqrt{12}} & D_s^- \\& & & \\
 D^0 & D^+ & D_s^+ & {-3\eta_c \over \sqrt{12}} \\ \end{array} \right) \label{phig}.
\end{eqnarray}

 A current is then defined:

\begin{equation}
{J}_\mu=[\dmu\Phi,\Phi]
\end{equation}
 and a Lagrangian is build by connecting two currents and adding an extra term proportional to the square mass of the fields:

\begin{eqnarray}
\lc_{PPPP}={1\over12f^2}Tr({J}_\mu {J}^\mu+\Phi^4 M) \label{lag}.
\end{eqnarray}
  $SU(4)$ and $SU(3)$ flavor symmetry 
 breaking already arise from the mass term when the matrix $M$ is not proportional to the identity
 matrix. We take:
 
 \begin{equation}
 M=\left( \begin{array}{cccc} m_\pi^2 &0&0&0 \\ 0&m_\pi^2&0&0 \\ 0&0&2m_K^2-m_\pi^2&0 \\
 0&0&0&2m_D^2-m_\pi^2 \end{array} \right).
 \end{equation}

 The term with the matrix $M$ is exactly the same one appearing in the chiral Lagrangian of eq. (\ref{chil}) after breaking $SU(4)$ and $SU(3)$ by means of (\ref{chisybr}). The term ${J}_\mu{J}^\mu$ in eq. (\ref{lag}) appears for four meson fields from the kinetic term, $\dmu U\dmuu U$, of the chiral Lagrangian in eq. (\ref{chil}) if $U$ is taken as a $SU(4)$ representation by means of replacing $\phi_8$ in eq. (\ref{phi8}) by $\Phi$ in eq. (\ref{phig})

 We will also implement a different sort of symmetry breaking in a way that we explain below.

 The constant $f$ appearing in the Lagrangian (\ref{lag}) is, in principle, the pion decay constant (in this work
 $f_\pi=93.0$ MeV). However, a different one will be used for the heavy mesons. In this latter case, the $f^2$ appearing
 in the amplitudes should be thought as the product of $\sqrt{f}$ for each meson leg in the corresponding vertex, with
 $f=f_\pi=93.0$ MeV for light mesons and $f=f_D=165$ MeV for heavy ones. This value for $f_D$ is of the order of magnitude expected from the experimental point of view [\ref{pdg}] and lattice calculations [\ref{lattice}].

 Directly applying Feynman rules to obtain transition amplitudes from this Lagrangian would be too much of a
 simplification. Indeed, the term ${J}_\mu{J}^\mu$ of the chiral Lagrangian is usually visualized as the exchange of a vector meson between pairs of pseudoscalar fields in the limit of $q^2<<m_V^2$ (the Weinberg Tomozawa term). In this case the kinetic term of the Lagrangian of eq. (\ref{lag}) is $SU(4)$ flavor symmetric and therefore implicitly assumes equal $m_V$ for all the exchanges of heavy and light vector mesons. In refs [\ref{lutz1}], [\ref{lutz2}], [\ref{chiang1}] an $SU(3)$ version of the interaction based on heavy quark symmetry is used which would correspond to allowing the exchange of only light vector mesons in the Weinberg-Tomozawa Lagrangian described by the derivative term of eq. (\ref{lag}), and neglecting the $M$ term [\ref{lutz3}]. In the present work we shall go one step further by allowing also the exchange of heavy vector mesons but weighted by their respective squared masses and we shall also keep the mass term as done in [\ref{flasybre1}], [\ref{flasybre2}], [\ref{walliser}]. In order to implement this we first decompose the $\Phi$ field into its $SU(3)$ components:

\begin{eqnarray}
 \Phi&=&\left( \begin{array}{cc}
 \phi_8+{1\over\sqrt{12}}\phi_1 \hat 1_3 & \phi_3 \\ \phi_{\bar3} & -{3\over\sqrt{12}}\phi_1
 \\ \end{array} \right) 
\end{eqnarray}

 The $\hat 1_3$ is the 3x3 identity matrix and the fields $\phi_i$ contain the meson fields for each $i$-plet of 
 $SU(3)$ into which the 15-plet of $SU(4)$ decomposes:

\begin{eqnarray*}
 \phi_8&=&\left( \begin{array}{ccc}
 {\pi^0 \over \sqd}+{\eta \over \sqs} & \pi^+ & K^+  \\
 \pi^- & {-\pi^0 \over \sqd}+{\eta \over \sqs} & K^0 \\
 K^- & \bar K^0 & {-2\eta \over \sqs} \\
  \end{array} \right) \\
 \phi_3&=&\left(\begin{array}{c} \bar D^0 \\ D^-\\ D_s^- \end{array} \right) \\
\phi_{\bar 3}&=&\left(\begin{array}{ccc} D^0 & D^+ & D_s^+ \end{array} \right) \\
 \phi_1&=& \eta_c
\end{eqnarray*}

 In this way the Lagrangian in (\ref{lag}) can be decomposed into six parts:

\begin{eqnarray}
 \lc_{PPPP}&=&{1 \over 12 f^2}(\lc_8+\lc_3+\lc_{31}+\lc_{83}+\lc_{831}+\lc_{mass}) \\
 \lc_8&=&Tr\left(J_{88_\mu} J_{88}^\mu\right) \\
 \lc_3&=&J_{\bar 3 3_\mu} J_{\bar 3 3}^\mu+Tr\left(J_{3 \bar 3_\mu} J_{3 \bar 3}^\mu \right) \label{lag3} \\
 \lc_{31}&=&{8\over3}J_{\bar 3 1\mu}J_{1 3}^\mu\\
 \lc_{83}&=&2\left(J_{\bar 3 8\mu}J_{8 3}^\mu+Tr\left(J_{\bar 3 3\mu}J_{88}^\mu\right)\right) \label{lag83}\\
 \lc_{831}&=&{4\over\sqrt{3}}\left(J_{\bar 3 1\mu}J_{8 3}^\mu+J_{\bar 3 8\mu}J_{1 3}^\mu\right)\\
 \lc_{mass}&=&Tr\left(M \Phi^4 \right)
\end{eqnarray}
where the currents are defined as $J_{ij}^\mu=(\dmuu\phi_i)\phi_j-\phi_i(\dmuu\phi_j)$.

 Now the exchange of charmed (heavy) vector mesons can be easily identified in the different pieces of the Lagrangian by identifying currents carrying explicitly charm quantum number. The
 $\lc_8$ term accounts for the exchange of light vector mesons only. In $\lc_{83}$ the first term is
 mediated by heavy vector mesons and the second term by light ones, $\lc_{831}$ and $\lc_{31}$ have only contributions from
 heavy vector mesons and $\lc_3$ will still have to be worked out further.

 The separation of the heavy vector meson contribution from $\lc_3$ is more subtle because the exchange of a heavy hidden charm meson in this 
 sector occurs in charge and flavor conserving hadronic currents, where the $\rho_0$ and $\omega$ also contribute. The strategy followed here
 is to construct a Lagrangian connecting the current ${J}_\mu$ with a vector field $V^\mu$ [\ref{lutz3}], [\ref{palomar}]:

\begin{eqnarray}
 \lc_{PPV}&=&-{i g \over \sqrt{2}}Tr\left([\dmu\Phi,\Phi]V^\mu\right) \label{lagppv}.
\end{eqnarray}
 Here $V^\mu$ is a 4x4 matrix with the same structure as $\Phi$, but with the 15-plet of vector mesons
 instead. The heavy vector meson which can be
 exchanged in charge and flavor conserving hadronic currents is the $J_\psi$. 

 The $J_\psi$ contribution can be calculated from the Lagrangian (\ref{lagppv}) and it is easy to see that when the vector mesons are connecting equal hadronic currents one has a contribution with weights $1\over3$ and $2\over3$ for light vector mesons and the $J_\psi$, respectively, while the weights are $-{1\over3}$ and $4\over3$ in terms connecting different currents. Appendix C shows in detail how to work out $\lc_3$.

 With all these considerations the full Lagrangian can now be rewritten in terms of the correction parameters:

\begin{eqnarray}
 \gamma&=&\left({m_L\over m_H}\right)^2 \label{gamafac} \\ 
 \psi_3&=&{1\over 3}+{2\over 3}\left({m_L\over m_{J_\psi}}\right)^2  \label{psi3fac} \\ 
 \psi_5&=&-{1\over 3}+{4\over 3}\left({m_L\over m_{J_\psi}}\right)^2 \label{psi5fac}.
\end{eqnarray}
 Here $m_L$ and $m_H$ are masses of light and heavy vector mesons. They will be set to $m_L=800$ MeV and $m_H=2000$ MeV. With these ingredients the full
 corrected Lagrangian can be written as:

\begin{eqnarray}
\lc&=&{1\over 12 f^2}\biggr(Tr\Big(J_{88\mu}J_{88}^\mu+2 J_{\bar 3 3\mu}J_{88}^\mu+J_{3 \bar 3_\mu} J_{3 \bar 3}^\mu\Big)+{8\over3}\gamma J_{\bar 3 1\mu}J_{1 3}^\mu+\nonumber\\
&&{4\over\sqrt{3}}\gamma\Big(J_{\bar 3 1\mu}J_{8 3}^\mu+J_{\bar 3 8\mu}J_{1 3}^\mu\Big)+2\gamma J_{\bar 3 8\mu}J_{8 3}^\mu+\psi_5 J_{\bar 3 3_\mu} J_{\bar 3 3}^\mu+\lc_{mass}\biggr)  \label{lagfull}.
\end{eqnarray}

 Note that from eq. (\ref{lagfull}) we can recover the usual lowest order chiral Lagrangian for $SU(3)$, which is the term proportional to $Tr(J_{88\mu}J_{88}^\mu)$, while the Lagrangian used by Kolomeitsev [\ref{lutz1}] and Guo [\ref{chiang1}], based on heavy quark symmetry [\ref{refe1}], [\ref{refe2}], [\ref{refe3}], [\ref{refe4}], is proportional to the term $Tr(J_{\bar 3 3\mu}J_{88}^\mu)$. Our model has also terms for the interaction of heavy mesons only, proportional to $Tr(J_{3 \bar 3_\mu} J_{3 \bar 3}^\mu)$ and $J_{\bar 3 3_\mu} J_{\bar 3 3}^\mu$ and all the other terms are corrections that can be controlled by the parameter $\gamma$.

 From this Lagrangian, applying the usual Feynman rules, the transition amplitudes in appendix A are calculated and
 used as potential for each possible reaction. This potential will be used as the kernel for
 solving the scattering equation.

 In order to support our results also the chiral Lagrangian with the flavor symmetry breaking pieces will be used to solve the scattering problem in the open charm sector. Very similar results are found and will be discussed in section 5.

\section{The Scattering Problem}
 
 The amplitudes needed, ${\cal M}(s,\theta)$, are written in Appendix A for the Lagrangian of eq. (\ref{lagfull}). Since we are only interested in s-wave meson-meson scattering, we first project the amplitudes over s-waves, by making a simple angular integration. After projecting the amplitudes for s-wave they will be transformed to isospin basis and inserted into the Bethe-Salpeter equation which in the on-shell formalism of [\ref{meme1}], [\ref{osetkn}] is reduced to an algebraic equation:
 
 \begin{equation}
 T=V+VGT \label{bseq}.
 \end{equation}
  In this equation $V$ is the potential, a matrix constructed with the tree level transition
 amplitudes for each one of the possible channels, projected over s-wave. The matrix G is diagonal
 with each one of its non-zero elements given by the loop function for the two particles in each channel:

\begin{eqnarray} 
 G_{ii}=&i\int {dq^4\over (2\pi)^4} {1\over q^2-m_1^2+i\epsilon}{1\over (P-q)^2-m_2^2+i\epsilon}=
 \label{loopint}\\
 &{1 \over 16\pi ^2}\biggr( \alpha _i+Log{m_1^2 \over \mu ^2}+{m_2^2-m_1^2+s\over 2s}
  Log{m_2^2 \over m_1^2}+\nonumber\\ 
  &{p\over \sqrt{s}}\Big( Log{s-m_2^2+m_1^2+2p\sqrt{s} \over -s+m_2^2-m_1^2+
  2p\sqrt{s}}+Log{s+m_2^2-m_1^2+2p\sqrt{s} \over -s-m_2^2+m_1^2+  2p\sqrt{s}}\Big)\biggr)
  \label{loopf}.
\end{eqnarray}
  P in equation (\ref{loopint}) is the total four-momentum of the two mesons in channel $i$ and $m_1$ and $m_2$ are
 the masses of the two mesons in this channel. The expression in eq. (\ref{loopf}) is calculated using dimensional regularisation.
  Over the real axis $p$ is the three-momentum of the mesons in the center of mass frame:

\begin{eqnarray}
p&=&{\sqrt{(s-(m_1+m_2)^2)(s-(m_1-m_2)^2)}\over 2\sqrt{s}} \label{trimom}.
\end{eqnarray}

  In the complex plane the momentum $p$ is calculated using the same expression. Eq. (\ref{bseq}) with eqs. (\ref{loopint}-\ref{loopf}) makes implicit use of dispersion relations in which only the right hand (physical) cut is considered. It was proved in [\ref{noverd}] that the left hand cut provides a moderate contribution, and more important, very weakly energy dependent, such that its contribution can be easily accommodated in terms of the subtraction constant that we use, in the range of energies of interest to us.

 In this work we will set the loop parameter in eq. (\ref{loopf}) to $\mu=1500$ MeV and fit the substraction
 constant, $\alpha$, as a free parameter.

  This loop function has the right imaginary part to ensure the unitarity of the T-matrix [\ref{meme2}]:

\begin{equation}
Im(G_{ii})=-{p\over 8\pi\sqrt{s}} \label{imloop}.
\end{equation}

  Equation (\ref{bseq}) can be easily inverted:

\begin{equation}
T=(\hat 1-VG)^{-1}V.
\end{equation}

 When looking for poles in the complex plane one should be careful because of the cuts of the loop function 
 beyond each threshold. Bound states appear as poles over the real axis and below threshold in the
 first Riemann sheet. Resonances show themselves as poles above threshold and in the second
 Riemann sheet of the channels which are open.

 Over the real axis the discontinuity of the loop function is known to be two times its imaginary part [\ref{inoue}]
 so, knowing the value of the imaginary part of the loop function over the axis, eq. (\ref{imloop}), one can do a proper analytic continuation of it for the whole complex plane:

\begin{eqnarray}
G_{ii}^{II}&=&G^{I}_{ii}+i {p\over 4\pi\sqrt{s}}, \hspace{1cm} Im(p)>0.
\end{eqnarray}
 $G^{II}$ and $G^I$ refer to the loop function in the second and first Riemanian sheets, respectively.

\begin{figure}
\begin{tabular}{cc}
\includegraphics[scale=0.3,angle=-90]{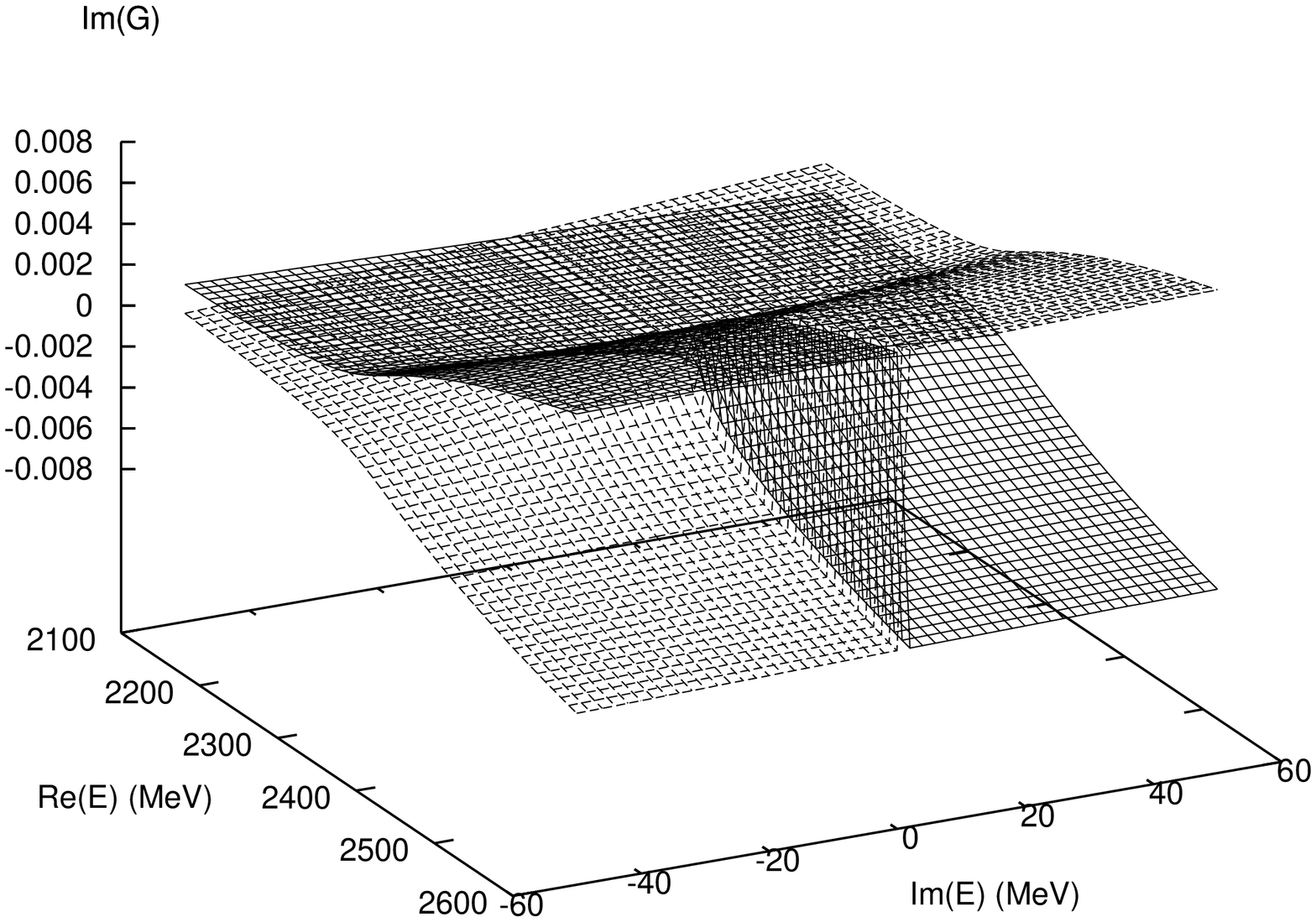} &
\includegraphics[scale=0.3,angle=-90]{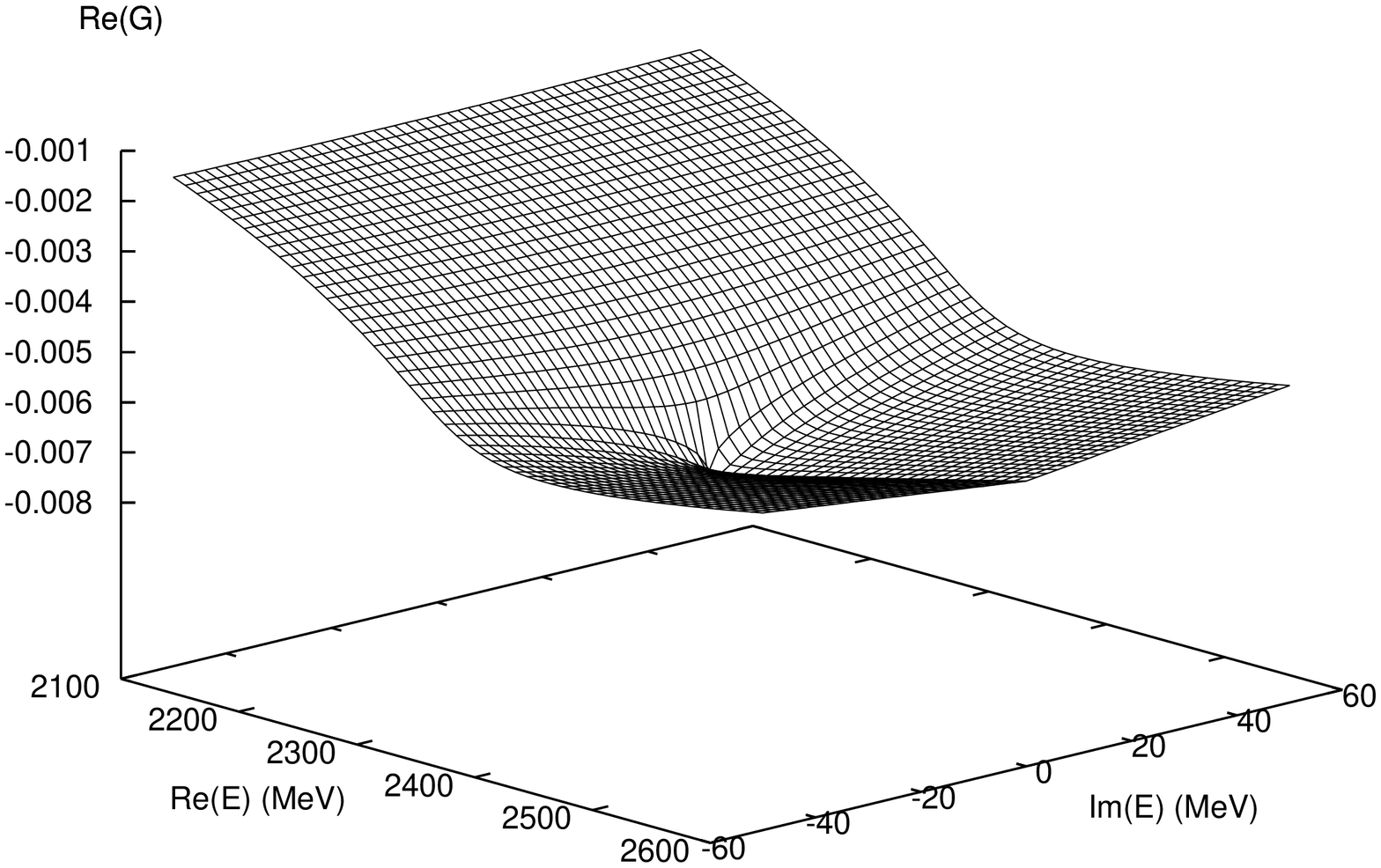} \\
\includegraphics[scale=0.3,angle=-90]{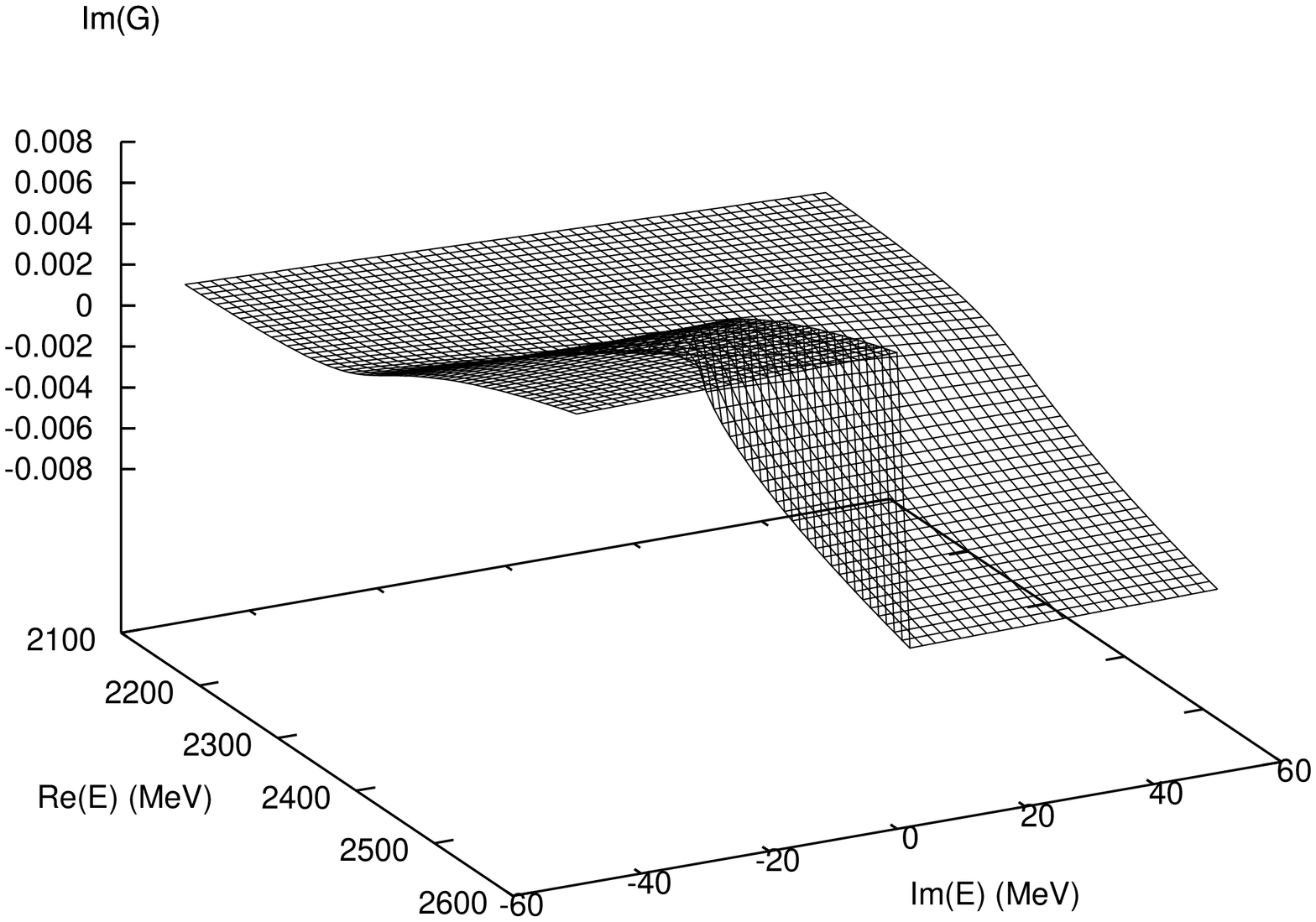} &
\includegraphics[scale=0.3,angle=-90]{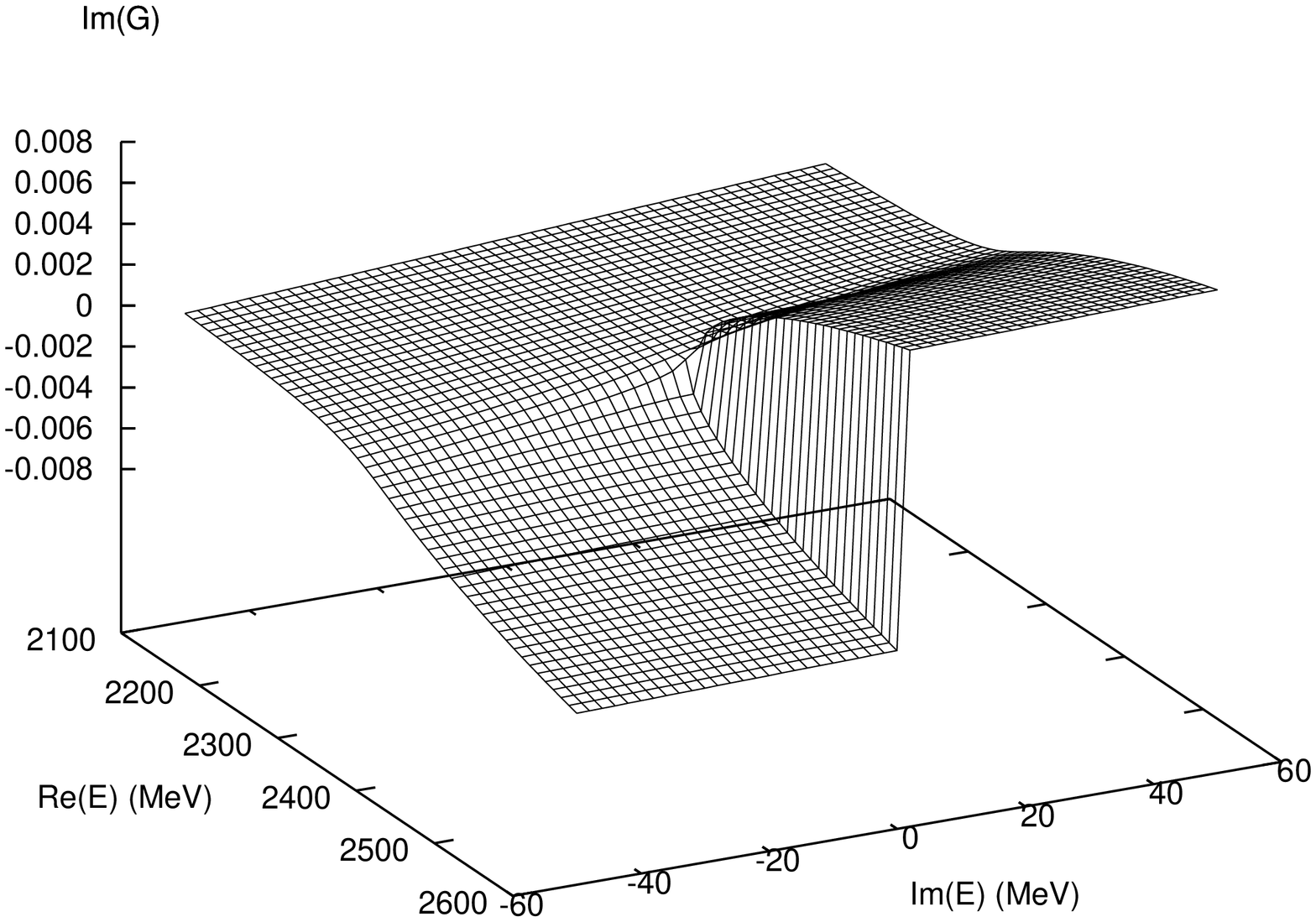} 
\end{tabular}
\caption{Upper left: Imaginary part of the loop on the first and second Riemann sheets superposed. Upper Right: Real part of
 the loop in the first Riemann sheet. Botton left, right are the imaginary part of the loop in the first and second
 Riemann sheets, respectively} \label{loop3d}
\end{figure}

 Figure \ref{loop3d} shows some plots of the loop function in the complex plane.

\section{Results}

 The amplitudes listed in appendix A are in a charge basis. First, they are transformed to an isospin basis and then
 to a $SU(3)$
 basis by means of the isospin and $SU(3)$ states given in appendix B. The $SU(3)$ symmetry breaking is then studied.
 The
 amplitudes in a $SU(3)$ basis show in which sectors the interaction is attractive (where it may generate resonances). The
 most important term in each amplitude is the $s$ term, so when an amplitude has a negative factor multiplying $s$
 it is considered to be attractive. The results of the diagonalization of the interaction in $SU(3)$ basis are as follows:

$8\otimes 8\rightarrow1\oplus8_s \oplus8_a\oplus10\oplus\bar{10}\oplus27$: The interaction here is repulsive in the 27; there is no interaction in the 10 nor in the $\bar{10}$; in one of the octets, because of its symmetry properties (here we should consider scattering of identical particles if SU(3) symmetry is restored) there is no interaction in even $l$
partial waves; in the other octet and in the singlet there is attraction, which will lead to the formation of 4 states, to be identified as the light scalar resonances, $\sigma$, $f_0$, $a_0$ and $\kappa$.

$\bar 3\otimes\bar 3\rightarrow3\oplus\bar 6$: Here there is no interaction in the 3 when the correction factors are set to 1, otherwise the interaction has a p-wave structure $(t-u)$. In the sextet the interaction is repulsive, therefore, no double charmed scalar resonances are expected from our model.

$\bar 3 \otimes 3\rightarrow8\oplus 1$: The interaction is attractive in both the octet and the singlet if the
 correction parameters ($\psi_3$, $\psi_5$) are set to 1. In this case, where the large mass of the $J_\psi$ is disregarded, one can see resonances generated. However, since the terms with the heavy vector meson have the largest weight in the amplitude, when the correction parameters are considered to take into account the different masses of the exchanged vector mesons, the resonances disappear for the octet. The singlet is always attractive irrespective to the correction parameters.

$\bar 3 \otimes 8\rightarrow\bar3\oplus6\oplus\bar{15}$: In the anti-triplet and sextet there is attraction while in the $\bar {15}$-plet there is repulsion. We generate in our model five resonances with charmed quantum number, two from the antitriplet and three from the sextet.

 We discuss below the free parameters of the theoretical framework and how we fit them. The parameters changed are $\alpha_H$ and $\alpha_L$. The $\alpha$'s are the subtraction constants for the loop functions. The parameter $\alpha_H$ was chosen for channels involving at least one heavy pseudoscalar meson and a different one for channels where there are just light ones, $\alpha_L$.

 One of the novel aspects of the present work is that we allow the mixing of the light mesons with the heavy ones in the search of zero charm or hidden charm scalar mesons. The first interesting result is that the influence of the heavy meson sector in the generation of the light scalar resonances ($\sigma$, $f_0$, $a_0$, $\kappa$) is negligible. For instance it was checked that different values $\alpha_H$ have very small effect over the pole position for the light resonances. Varying $\alpha_H$ between -0.3 and -2.3 has less than 10\% effect over the pole position of the $f_0$ resonance, for example. So the heavy sector can be worked independently from the light one.

 With this in mind the open-charm (C=1) sector was used to fit $\alpha_H$ so that the position of the pole in the S=1, I=0 sector match the $D_{s0}^*(2317)$, which has already been suggested as being dynamically generated in [\ref{lutz1}], [\ref{lutz2}] and [\ref{chiang1}]. After fixing the heavy parameter, the $\alpha_L$ was fitted by locating the pole position in the sector C=0, S=0, I=1, which correspond to the $a_0$ resonance. We also made the fit of $\alpha_H$ for the model involving the chiral Lagrangian. The results are as follows:

 Phenomenological model: $\alpha_H$=-1.3 and $\alpha_L$=-1.3

 Chiral model: $\alpha_H$=-1.15 (we only applied this model for the open charm sector.)

\subsection{$SU(3)$ Symmetry Breaking}

 In our phenomenological model it is assumed that the $SU(3)$ flavor symmetry breaking arises from the different masses of the interacting mesons. The mass used for each member of the 15-plet is:

 $m_\pi$=138.0 MeV, $m_K$=495.0 MeV, $m_\eta$=548.0 MeV, 

 $m_D$=1865.0 MeV, $m_{D_s}$=1968.0 MeV and $m_{\eta_c}$=2979.0 MeV

 Note that there is no isospin breaking in the model, all particles in a same isospin multiplet are considered to have
 the same mass. So, in this work, the Bethe-Salpeter equation, eq. (\ref{bseq}), was solved with $V$ in isospin basis.

 $SU(3)$ symmetry can then be gradually broken by means of a symmetry breaking parameter $x$ which takes values between 1 and
 0, 1 meaning symmetry broken as we see it in the real world, and 0 symmetry restored. The masses of the mesons as a 
 function of the parameter $x$ are given by:

\begin{eqnarray}
m(x)=\bar m+x(m_{phys.}-\bar m)
\end{eqnarray}
where $\bar m$ is the meson mass in the symmetry limit.

 Two different values of $\bar m$ were used, for the light mesons (the ones belonging to the octet) it was set to 
 430 MeV and for the heavy ones, 1900 MeV.

 Also the correction parameters were changed along with $x$, although they just violate $SU(4)$ symmetry:

\begin{eqnarray}
\gamma(x)=1+x(\gamma_{phys.}-1)
\end{eqnarray}

 Similar functions are constructed for $\psi_3$ and $\psi_5$.

\begin{figure}
\begin{center}
\includegraphics[scale=1.0,angle=-0]{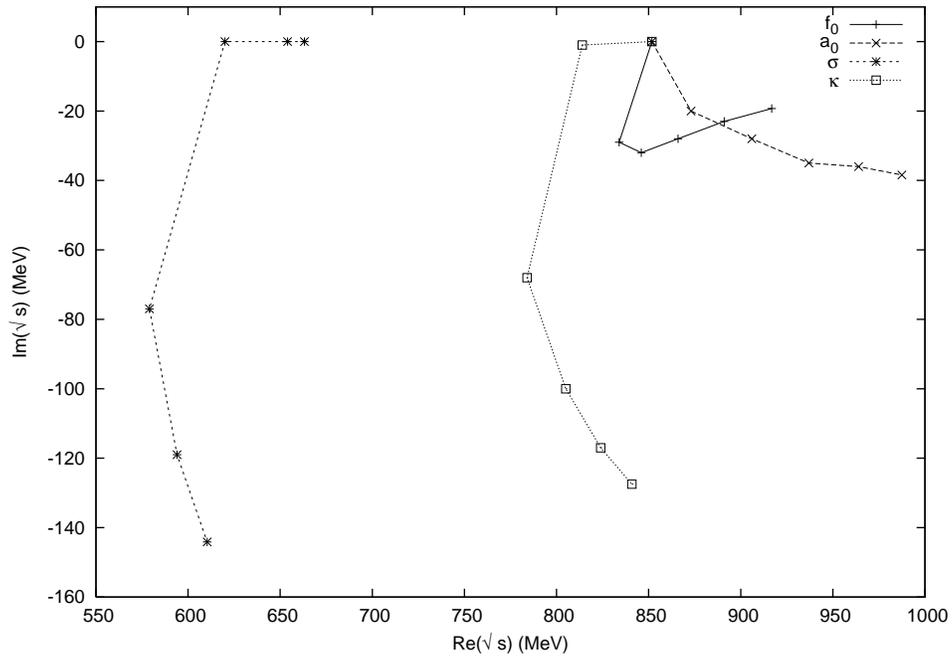} 
\caption{Light Scalars. The octet starts from E=851.76 MeV and the singlet from 663.13 MeV.} \label{c0l}
\end{center}
\end{figure}

\begin{figure}
\begin{center}
\includegraphics[scale=1.0,angle=-0]{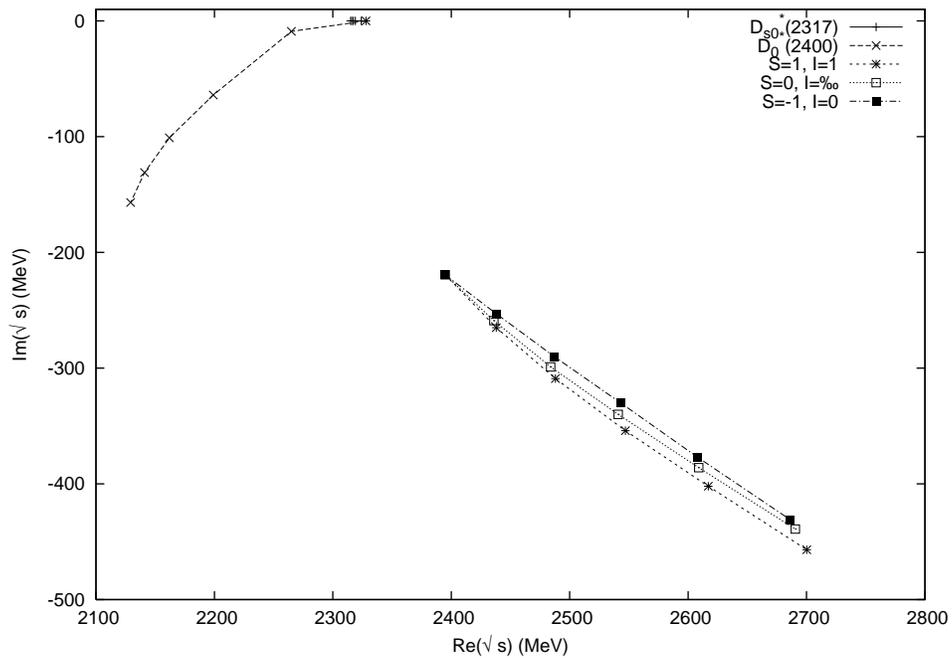} 
\caption{Heavy Scalars. The antitriplet starts from 2327.96 MeV and the sextet from (2394.87-$i $ 219.33)MeV} \label{c1h}
\end{center}
\end{figure}

 All scalar resonances in the same multiplet have the same mass once $SU(3)$ is restored while its breaking
 splits the masses of the different isospin multiplets. So, when written in the $SU(3)$ basis, the non-diagonal elements of the matrix $V$ (the ones which represent mixing between different $SU(3)$ multiplets) are always proportional to $m_\pi^2-m_K^2$.  Figures \ref{c0l} and \ref{c1h} show the results for the
 C=0 and C=1 sectors, varying $x$ from 0 to 1 in steps of 0.2. We should note that some resonances, for example, the $\kappa$ and the $D_0^*(2400)$ appear as cusps for small $x$. This happens because thresholds appear during the symmetry breaking procedure.

 Table \ref{poleposexp} displays the experimental situation of the scalar resonances, Table \ref{poleposset1} shows the results in the open charm sector for the problem solved with the chiral Lagrangian and Table \ref{poleposset2} shows the pole positions found within the phenomenological model developed in this work.

 In the following the results for each sector will be discussed separately. 

\begin{table}[h]
\begin{center}
\begin{tabular}{c|c|c|c|c|c}
\hline
Resonance ID& C&S&I& Mass (MeV)& $\Gamma$ (MeV)\\
\hline
\hline
$f_0$&0&0&0&980$\pm$10&40-100\\
\hline
$\sigma$&0&0&0&400-1200&250-500\\
\hline
$a_0$&0&0&1&984.7$\pm$1.2&50-100\\
\hline
$\kappa$&0&1&$1\over2$&841$\pm$30$^{+81}_{-73}$&618$\pm$90$^{+96}_{-144}$\\
\hline
$D_{s0}^*(2317)$&1&1&0&2317.3$\pm$0.4$\pm$0.8& $<$ 4.6\\
\hline
$D_0^*(2400)$&1&0&$1\over2$&2403$\pm$14$\pm$35&283$\pm$24$\pm$34\\
             & & &                       &2352$\pm$50       &261$\pm$50\\
\hline
\end{tabular}
\caption{Data from [\ref{pdg}]} \label{poleposexp}
\end{center}
\end{table}

\begin{table}[h]
\begin{center}
\begin{tabular}{c|c|c|c|c|c}
\hline
Resonance ID& C&S&I& RE($\sqrt{s}$) (MeV)& IM($\sqrt{s}$) (MeV)\\
\hline
\hline
$D_{s0}^*(2317)$&1&1&0&2315.41& 0\\
\hline
$D_0^*(2400)$&1&0&$1\over2$&2147.65&-107.29\\
\hline
(?)&1&0&$1\over2$&No pole &- \\
\hline
(?)&1&1&1&2427.70 &-248.40 \\
\hline
(?)&1&-1&0&2410.26 &-193.80 \\
\hline
\end{tabular}
\caption{Pole positions for the chiral Lagrangian} \label{poleposset1}
\end{center}
\end{table}

\begin{table}[h]
\begin{center}
\begin{tabular}{c|c|c|c|c|c}
\hline
Resonance ID& C&S&I& RE($\sqrt{s}$) (MeV)& IM($\sqrt{s}$) (MeV)\\
\hline
\hline
$f_0$&0&0&0&916.83&-19.28\\
\hline
$\sigma$&0&0&0&610.33&-144.10\\
\hline
(?)&0&0&0&3718.87&-0.11\\
\hline
$a_0$&0&0&1&987.45&-38.40\\
\hline
$\kappa$&0&1&$1\over2$&840.91&-127.48\\
\hline
$D_{s0}^*(2317)$&1&1&0&2317.32& 0\\
\hline
$D_0^*(2400)$&1&0&$1\over2$&2129.22&-156.95\\
\hline
(?)&1&0&$1\over2$&2690.50 &-439.09 \\
\hline
(?)&1&1&1&2700.15 &-456.88 \\
\hline
(?)&1&-1&0&2686.13 &-430.87 \\
\hline
\end{tabular}
\caption{Pole positions for the phenomenological model} \label{poleposset2}
\end{center}
\end{table}

\subsection{C=0, S=0, I=0}

 Our model successfully generates poles which can be associated with the known light scalar resonances. In this sector, in the low energy region, 
 two poles can be found in the T-matrix, one corresponding to the $f_0$, but with a lower mass than one expects and
 another one for the $\sigma$. It is actually possible to adjust the mass of the $f_0$ pole in our model by increasing the $\alpha_L$ parameter, but two prices are paid: first the $a_0$ pole in the $S=0$, $I=1$ sector disappears for much bigger $\alpha_L$ and also the width of a more massive $f_0$ decreases. 

 Two more poles can be expected in this sector, one from the octet and the other one from the singlet, coming from the scattering of the heavy
 mesons. For $x$=0 both poles appear, the singlet always very narrow, because its coupling to the light channels is very suppressed, and the octet with a much bigger width. The octet state disappears before $x=1.0$. One should notice that the width found for this new heavy resonance is very small. This happens because, as mentioned (see Table \ref{s0i0res}), the couplings to the light channels are very suppressed and the other possible decay channel is an octet formed by $\eta_c$ with a light meson which violates $SU(3)$. Table \ref{s0i0res} shows the absolute value of the residues for the resonances in this sector.

\begin{table}[h]
\begin{center}
\begin{tabular}{c|c|c|c}
\hline
Channel&$f_0$ res (GeV)&$\sigma$ res (GeV)& Heavy Singlet res (GeV)\\
\hline
\hline
& & &\\
$\pi\pi$&1.96&4.23&0.21\\
\hline
& & &\\
$K\bar{K}$&3.82&1.28&0.03\\
\hline
& & & \\
$\eta\eta$&4.47&0.47&0.00\\
\hline
& & &\\
$D\bar{D}$&0.71&4.08&10.41\\
\hline
& & &\\
$D_s\bar{D_s}$&3.73&0.49&6.73\\
\hline
& & &\\
$\eta\eta_c$&2.07&1.04&0.29\\
\hline
\end{tabular}
\caption{Residues for the poles in the C=0, S=0, I=0 sector} \label{s0i0res}
\end{center}
\end{table}

\begin{figure}
\begin{center}
\includegraphics[scale=1.0,angle=0]{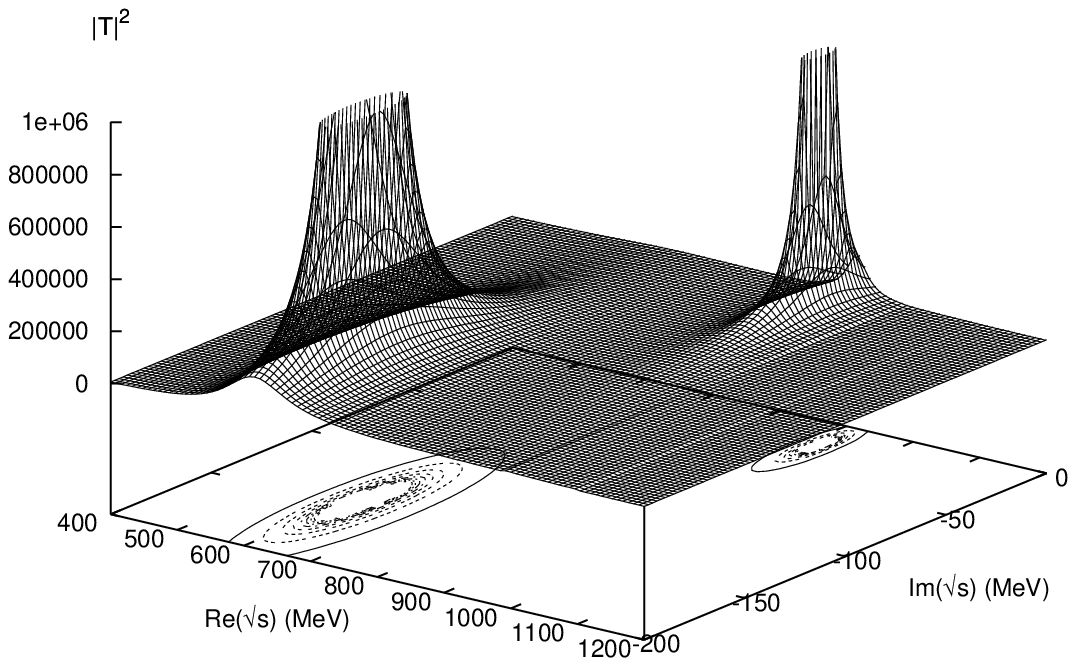} 
\caption{$T T^\dagger$ for $\pi\pi$-channel in C=0, S=0, I=0 sector} \label{poles}
\end{center}
\end{figure}

 Figure \ref{poles} shows the absolute value of the square of the transition matrix for this sector, as an
 illustration.

\subsection{C=0, S=0, I=1}

 In this sector the model successfully generates the $a_0$ resonance. Both the mass and the width found for it in the model 
 agree very well with experimental values. Note, however that this sector was actually used to fit $\alpha_L$, but fitting just this one parameter, both the width and the mass for the $a_0$ are in good agreement with experiment. As mentioned if we used the pole position of the $f_0$ resonance to adjust the parameter $\alpha_L$ we would lose the $a_0$ pole. This relative instability of the $a_0$ resonance with respect to the parameters of the theory is not new, it also occurs when using the inverse amplitude method for unitarization and the potential of the lowest order chiral Lagrangian where the $a_0$ appears as a cusp and not a pole. The pole is, however, regained when the information of the second order Lagrangian is used as input in the potential [\ref{meme2}].

 Table \ref{s0i1res} shows its couplings\footnote{because of the identical particles, the $\pi\pi$ channel in $I=1$ just contributes to odd parity partial waves (indeed, the amplitude has a p-wave
 structure $t-u$).} to the different channels.

 In the heavy sector again the pole for the octet just appears for small values of $x$.

\begin{table}[h]
\begin{center}
\begin{tabular}{c|c}
\hline
Channel&$a_0$ res (GeV)\\
\hline
\hline
&  \\
$\pi\pi$&0\\
\hline
&  \\
$K\bar{K}$&3.84\\
\hline
&  \\
$\eta\pi$&2.65\\
\hline
&  \\
$D\bar{D}$&3.64\\
\hline
& \\
$\pi\eta_c$&1.63\\
\hline
\end{tabular}
\caption{Residues for the $a_0$ pole} \label{s0i1res}
\end{center}
\end{table}

\subsection{C=0, S=1, I=${1\over2}$}

 The pole generated here should be identified with the $\kappa$ resonance. This resonance, however, is a very broad one
 and although there is debate on the existence of this resonance, recent experiments have come to support it [\ref{pes}]-[\ref{pes6}].

 Again there are no heavy resonances for $x$=1.

\begin{table}[h]
\begin{center}
\begin{tabular}{c|c}
\hline
Channel&$\kappa$ res (GeV)\\
\hline
\hline
&  \\
$K\pi$&3.81\\
\hline
&  \\
$K\eta$&2.00\\
\hline
&  \\
$D_s\bar{D}$&4.10\\
\hline
&  \\
$K\eta_c$&2.25\\
\hline
\end{tabular}
\caption{Residues for the $\kappa$ pole} \label{kappa}
\end{center}
\end{table}

\subsection{C=1, S=1, I=0}

 The $D_{s0}^*(2317)$ is reproduced in this work as a mixed bound state of $|DK>$ and $|D_s\eta>$. Experimentally the 
 observed decay channel for this resonance is $D_s\pi$ which is not allowed in the model because it is an isospin violating process.
 However if one considers isospin violation by solving the Bethe-Salpeter equation for charge eigenstates instead of 
 isospin ones and considering the real masses of the mesons, including the differences between different $I_3$
 components,
 one gets a very narrow width of less then a keV for this resonance. Another possible source of contribution is to consider $\eta-\pi^0$ mixing by means of which in [\ref{chiang1}] one gets a width of the order of a few keV. The width of this resonance is given as an upper bound 
 of about 4 MeV in [\ref{pdg}].

 The couplings of this pole to the channels is shown in Table \ref{ds02317} for both models considered in this work.

\begin{table}[h]
\begin{center}
\begin{tabular}{c|c|c}
\hline
Channel& Chiral & Phenom.\\
      & model res (GeV)& model res (GeV)\\
\hline
\hline
& & \\
$DK$&10.21&10.36\\
\hline
& & \\
$D_s\eta$&6.40&6.00\\
\hline
& & \\
$D_s\eta_c$&0.48&1.52\\
\hline
\end{tabular}
\caption{Residues for the $D_{s0}^*(2317)$ pole} \label{ds02317}
\end{center}
\end{table}

\subsection{C=1, S=0, I=${1\over2}$}

 Two poles are found here, one is the antitriplet companion of the $D_{s0}^*(2317)$, also experimentally known and to
 be identified as $D_0^*(2400)$. Although the antitriplet pole generated by the model in this sector has a width in 
 agreement with the experimental value, the model fails in predicting its mass by around 150 MeV, which might not be too serious considering that the experimental width is around 300 MeV.

 Additionally another state is generated, belonging to the sextet. Here the two models differ from each other. In the chiral model this resonance has a smaller mass and width, but disappears as $x$ reaches 1 because of thresholds effects, while the pole is predicted by our model around 2700 MeV but with a huge width that makes it irrelevant from the experimental point of view.

 Residues for the $D_0^*(2400)$ pole are in Table \ref{d02400} for both models.

\begin{table}[h]
\begin{center}
\begin{tabular}{c|c|c}
\hline
Channel& Chiral & Phenom.\\
      & model res (GeV)& model res (GeV)\\
\hline
\hline
& \\
$D\pi$&8.91&10.87\\
\hline
& \\
$D\eta$&1.36&3.77\\
\hline
&  \\
$D_s\bar{K}$&5.71&8.52\\
\hline
& \\
$D\eta_c$&3.20&2.14\\
\hline
\end{tabular}
\caption{Residues for the $D_0^*(2400)$ pole.} \label{d02400}
\end{center}
\end{table}

\subsection{C=1, S=1, I=1 and C=1, S=-1, I=0}

 The other two states belonging to the sextet are to be found in these sectors. However they differ in mass and width from one model to the other. While with the chiral Lagrangian these poles have mass around 2400 MeV and width about 0.5 GeV, within our model their mass is 300 MeV larger and the huge width of the order of 1 GeV would make these poles irrelevant from the experimental point of view.

\subsection{Comparison With Other Works}

 The light scalar resonances reproduced in this work have been thoroughly investigated in more sophisticated approaches and with higher orders of the chiral Lagrangian [\ref{meme1}], [\ref{meme2}], [\ref{meme3}], [\ref{refee1}], [\ref{refee2}]. In our study of the hidden charm states we have now used coupled channels involving light and heavy pseudoscalar mesons and we find actually a negligible mixing of the two sectors.

 The open charm sector has been studied by Kolomeitsev [\ref{lutz1}] and Guo [\ref{chiang1}] in a very similar framework but with different Lagrangians from ours; both have used the same Lagrangian, and very similar parameters. In [\ref{lutz2}] higher order chiral Lagrangians are used in this sector. The Lagrangian in these works neglects exchange of heavy vector mesons while the present work includes it although suppressed in a proper way. The second term of the Lagrangian in eq. (\ref{lag83}) can be identified with the lowest order chiral Lagrangian used in [\ref{lutz1}] and [\ref{chiang1}] except that in the present work this term of the Lagrangian is a factor $3\over2$ smaller. Another difference between this present work and previous ones is the meson decay constant, $f$. In previous works it was always set to the pion decay constant, while in the present one, inspired by experimental measurements and lattice calculations we use a different value for the decay constant of the charmed mesons. 

 In the S=1, I=0 sector the results of all works coincide and the $D_{s0}^*(2317)$ is well reproduced. Its anti-triplet companion, the $D_0^*(2400)$, is also well reproduced in the S=0, I=$\oh$ sector. However, in this sector the present work differs from previous ones: while within our model, the sextet state is extremely broad, in the works of Kolomeitsev and Guo a narrow state is predicted in this sector. The chiral Lagrangian we used seams to give an intermediate situation between our work and these previous ones, it generates for the sextet states a broad resonance although not as broad as in our model. The huge width of these resonances within our model is also a consequence of its much bigger mass which causes a much bigger phase space for decaying into the open channels.

 Another novelty in the present work is the study of the hidden charm sector. Here we mixed light with heavy pseudoscalar pairs and concluded that there was barely any mixing of the heavy and light sectors. This result supports the findings for the light scalars, using only light pseudoscalar mesons as building blocks. On the other hand we find a heavy scalar with mass around 3.7 GeV corresponding mostly to a $D\bar D$ state, very narrow (see Table \ref{poleposset2}), which is surprising in view of the large phase space available for decay into pairs of pseudoscalars. The dynamics which prevents the mixing of the heavy and light sectors is responsible for this very small width.

 We should also note that with a different formalism using the Schr\"odinger equation with one vector-meson exchange potential, $D\bar D$ states also appear for some choices of a cut off parameter in [\ref{chiang2}].

\section{Summary}

We studied the dynamical generation of resonances in a unitarized coupled channel framework. We constructed a Lagrangian based on $SU(4)$ flavor symmetry and after decomposing the field of pseudoscalar mesons in this Lagrangian into its $SU(3)$ components, we were able to identify terms mediated by exchange of heavy vector mesons and thus suppress these, hence breaking the $SU(4)$ structure of the Lagrangian. The results were also compared with previous works based on chiral theory and heavy quark symmetry and with results obtained from a chiral Lagrangian considering flavor symmetry breaking effects.

 The amplitudes calculated from this Lagrangian, written in a $SU(3)$ basis, show in which sectors the interaction is attractive so that it might generate resonances. Within the framework developed in the present work an $SU(3)$ octet and a singlet of scalar mesons appear in the light sector. These resonances can be identified with the light scalar mesons, $\sigma$, $f_0$, $a_0$ and $\kappa$, which have been thoroughly investigated before, but which also show up, practically undisturbed, in the enlarged basis of coupled channels used in the present work.

 In the heavy sector an anti-triplet is generated leading to two states which can be identified with the controversial $D_{s0}^*(2317)$ and $D_0^*(2400)$ states, though the mass generated for this second one is somewhat lower than the experimental one. Thus, in the framework developed here, these scalar states should be interpreted as bound and quasi-bound states in coupled channels: The $D_{s0}^*(2317)$ being mainly a $DK$ bound state with no decay, except for a tiny one when allowing isospin violation and the $D_0^*(2400)$ a $D\pi$ resonance.

Also a very broad sextet is generated in the heavy sector, but these states are extremely broad, making them irrelevant from the experimental point of view. One should note, however that these broad states contrast with states generated in previous works [\ref{lutz1}], [\ref{chiang1}] where narrow structures are found with the same quantum numbers. The Lagrangian in these previous works neglects the exchange of heavy vector mesons and uses a much stronger coupling.

Also a heavy singlet appears as a pole in the T-matrix. This singlet comes from the attraction generated in the $\bar 3 \otimes 3$ sector and its structure is mainly a $D\bar D$ quasi-bound state.

\section{Acknowledgments}  This work is partly supported by DGICYT contract
number BFM2003-00856, the Generalitat Valenciana and the E.U. EURIDICE 
network
contract no.  HPRN-CT-2002-00311. This research is  part of the EU 
Integrated
Infrastructure Initiative  Hadron Physics Project under  contract number
RII3-CT-2004-506078.

\newpage

\appendix

\section{Amplitudes}

 This appendix shows the amplitudes obtained from the Lagrangian in eq. (\ref{lagfull}). In the column of the states, the following 
 momenta assignments should be taken into account: where it reads $M_1 M_2 \rightarrow M_3 M_4$ it means
 $M_1(p) M_2(k) \rightarrow M_3(p') M_4(k')$ and the Mandelstam variables are defined as follows:

\begin{eqnarray}
 s=&(p+k)^2=&(p'+k')^2 \\
 t=&(p-p')^2=&(k-k')^2 \\
 u=&(p-k')^2=&(k-p')^2 
\end{eqnarray}

 When inserting these amplitudes (or transformed to isospin or $SU(3)$ basis) in the BS-equation, one should be
 careful to divide the amplitude  by $\squd$ each time the initial or the final state contains a pair of identical
 particles, (unitary normalization) in order to ensure closure of the intermediate states. The extra normalization for the external lines must be kept in mind but does not matter for the pole search. The factors $\gamma$, $\psi_3$ and $\psi_5$ are defined in eqs. (\ref{gamafac})-(\ref{psi5fac}).

\subsection{C=2, S=2}

\begin{tabular}{|c|c|}
\hline
States  & Amplitude \\
\hline
\hline
$D_s^+D_s^+ \rightarrow D_s^+D_s^+$ &  $-{1\over3f^2} (-\psi_3(2s-t-u)+2m_D^2+2m_K^2-2m_\pi^2) $ \\
\hline
\end{tabular}

\subsection{C=2, S=1}

\begin{tabular}{|r|c|}
\hline
States  & Amplitude \\
\hline
\hline
$D_s^+D^0 \rightarrow D_s^+D^0$  & $-{1\over6f^2}(-\psi_5(s-u)-(s-t)+2m_D^2+m_K^2-m_\pi^2)$ \\
\hline
\end{tabular}

\subsection{C=2, S=0}

\begin{tabular}{|r|c|}
\hline
States  & Amplitude \\
\hline
\hline
$D^+D^0 \rightarrow D^+D^0$  & $-{1\over6f^2}(-\psi_5(s-u)-(s-t)+2m_D^2)$ \\
\hline
\end{tabular}

\subsection{C=1, S=2}

\begin{tabular}{|r|c|}
\hline
States & Amplitude \\
\hline
\hline
$K^0D_s^+ \rightarrow K^0 D_s^+$  & $-{1\over6f^2}(-(s-u)-\gamma(s-t)+m_D^2+2m_K^2-m_\pi^2)$ \\
\hline
\end{tabular}

\subsection{C=1, S=1}

\begin{tabular}{|r|c|}
\hline
States & Amplitude \\
\hline
\hline
$K^+D^0 \rightarrow K^+D^0$  & $-{1\over6f^2}(\gamma(t-u)+(s-u)+m_D^2+m_K^2)$ \\
   $\rightarrow K^0D^+$  & $-{1\over6f^2}(\gamma(t-u)+(s-u)+m_D^2+m_K^2)$ \\
   $\rightarrow \pi^0D_s^+$  & $-{1\over6\sqrt{2}f^2}(-(s-u)-\gamma(s-t)+m_D^2+m_K^2)$ \\
   $\rightarrow \eta D_s^+$ & $-{1\over6\sqrt{6}f^2}(\gamma(u-t)-(3+\gamma)(s-u)-m_D^2-3m_K^2+2m_\pi^2)$\\
\hline
$K^0D^+\rightarrow K^0D^+$  & $-{1\over6f^2}(\gamma(t-u)+(s-u)+m_D^2+m_K^2)$\\
 $\rightarrow \pi^0D_s^+$ & $-{1\over6\sqrt{2}f^2}((s-u)+\gamma(s-t)-m_D^2-m_K^2)$\\
 $\rightarrow \eta D_s^+$  & $-{1\over6\sqrt{6}f^2}(\gamma(u-t)-(3+\gamma)(s-u)-m_D^2-3m_K^2+2m_\pi^2)$\\
\hline
$\pi^0D_s^+\rightarrow \pi^0D_s^+$  & - \\
$\rightarrow \eta D_s^+$  & - \\
\hline
$\eta D_s^+\rightarrow \eta D_s^+$  & $-{1\over9f^2}(\gamma(-s+2t-u)+2m_D^2+6m_K^2-4m_\pi^2)$\\
\hline
$\eta_c D_s^+\rightarrow \eta_c D_s^+$  & $-{1\over18f^2}(4\gamma(-s+2t-u)+11m_D^2+3m_K^2-7m_\pi^2)$\\
 $\rightarrow K^+D^0$  & $-{1\over6\sqrt{3}f^2}(2\gamma(-s+2t-u)-2m_D^2+m_\pi^2)$\\
 $\rightarrow K^0D^+$  & $-{1\over6\sqrt{3}f^2}(2\gamma(-s+2t-u)-2m_D^2+m_\pi^2)$\\
 $\rightarrow \pi^0D_s^+$  & -\\
 $\rightarrow \eta D_s^+$  & $-{1\over9\sqrt{2}f^2}(-2\gamma(-s+2t-u)+2m_D^2-m_\pi^2)$\\
\hline
\end{tabular}

\subsection{C=1, S=0}

\begin{tabular}{|r|c|}
\hline
States  & Amplitude \\
\hline
\hline
$\pi^0D^0\rightarrow \pi^0D^0$ & $-{1\over12f^2}(\gamma(-s+2t-u)+2m_D^2+2m_\pi^2)$\\
$\rightarrow \pi^-D^+$  & $-{1\over6\sqrt{2}f^2}((2+\gamma)(s-u))$ \\
$\rightarrow \eta D^0$  & $-{1\over12\sqrt{3}f^2}(\gamma(-s+2t-u)+2m_D^2+2m_\pi^2)$\\
$\rightarrow K^- D_s^+$  & $-{1\over6\sqrt{2}f^2}(\gamma(t-u)+s-u+m_D^2+m_K^2)$\\
\hline
$\pi^-D^+\rightarrow \pi^-D^+$  & $-{1\over6f^2}(\gamma(t-u)+s-u+m_D^2+m_\pi^2)$\\
$\rightarrow \eta D^0$  & $-{1\over6\sqrt{6}f^2}(\gamma(-s+2t-u)+2m_D^2+2m_\pi^2)$\\
$\rightarrow K^-D_s^+$  & $-{1\over6f^2}(\gamma(t-u)+s-u+m_D^2+m_K^2)$\\
\hline
$\eta D^0\rightarrow \eta D^0$  & $-{1\over36f^2}(\gamma(-s+2t-u)+2m_D^2+2m_\pi^2)$\\
$\rightarrow K^-D_s^+$  & $-{1\over6\sqrt{6}f^2}(\gamma(s-t)+(3+\gamma)(s-u)-m_D^2-3m_K^2+2m_\pi^2)$\\
\hline
$K^-D_s^+\rightarrow K^-D_s^+$  & $-{1\over6f^2}(\gamma(t-u)+s-u+m_D^2+2m_K^2-m_\pi^2)$\\
\hline
$\eta_c D^0\rightarrow \eta_c D^0$  & $-{1\over18f^2}(4\gamma(-s+2t-u)+11m_D^2-4m_\pi^2)$\\
$\rightarrow \pi^0D^0$  & $-{1\over6\sqrt{6}f^2}(2\gamma(-s+2t-u)-2m_D^2+m_\pi^2)$\\
$\rightarrow \pi^-D^+$  & $-{1\over6\sqrt{3}f^2}(2\gamma(-s+2t-u)-2m_D^2+m_\pi^2)$\\
$\rightarrow \eta D^0$  & $-{1\over18\sqrt{2}f^2}(2\gamma(-s+2t-u)-2m_D^2+m_\pi^2)$\\
$\rightarrow K^-D_s^+$  & $-{1\over6\sqrt{3}f^2}(2\gamma(-s+2t-u)-2m_D^2+m_\pi^2)$\\
\hline
\end{tabular}

\subsection{C=1, S=-1}

\begin{tabular}{|r|c|}
\hline
States  & Amplitude \\
\hline
\hline
$K^-D^+ \rightarrow K^-D^+$  &- \\
$\rightarrow \bar K^0D^0$  & $-{1\over6f^2}(-(s-u)-\gamma(s-t)+m_D^2+m_K^2)$ \\
\hline
$\bar K^0D^0 \rightarrow \bar K^0D^0$  &- \\
\hline

\end{tabular}

\subsection{C=0, S=1}

\begin{tabular}{|r|c|}
\hline
States  & Amplitudes  \\
\hline
\hline
$D_s^+D^- \rightarrow D_s^+D^-$  & $-{1\over6f^2}(t-u+\psi_5(s-u)+2m_D^2+m_K^2-m_\pi^2)$ \\
\hline
$\pi^0K^0 \rightarrow \pi^0K^0$  & $-{1\over12f^2}(-s+2t-u+2m_K^2+2m_\pi^2)$ \\
$\rightarrow \pi^-K^+$ & $-{1\over2\sqrt{2}f^2}(-s+u)$ \\
$\rightarrow \eta K^0$ & $-{1\over12\sqrt{3}f^2}(3(s-2t+u)+2m_K^2-2m_\pi^2)$ \\
\hline
$\pi^-K^+\rightarrow \pi^-K^+$  & $-{1\over6f^2}(s+t-2u+m_K^2+m_\pi^2)$ \\
$\rightarrow \eta K^0$  & $-{1\over6\sqrt{6}f^2}(-3(s-2t+u)-2m_K^2+2m_\pi^2)$ \\
\hline
$\eta K^0\rightarrow \eta K^0$  & $-{1\over12f^2}(-3(s-2t+u)+6m_K^2-2m_\pi^2)$ \\
\hline
$\eta_c K^0\rightarrow \eta_c K^0$ & $-{1\over6f^2}m_K^2$ \\
\hline
$D_s^+D^- \rightarrow \pi^0K^0$  & $-{1\over6\sqrt{2}f^2}(t-u-\gamma(s-t)-m_D^2-m_K^2)$ \\
$\rightarrow \pi^-K^+$ & $-{1\over6f^2}(-(t-u)+\gamma(s-t)+m_D^2+m_K^2)$ \\
$\rightarrow \eta K^0$ & $-{1\over6\sqrt{6}f^2}((3+\gamma)(u-t)-\gamma(s-u)-m_D^2-3m_K^2+2m_\pi^2)$ \\
\hline
$D_s^+D^- \rightarrow \eta_c K^0$  & $-{1\over6\sqrt{3}f^2}(2\gamma(2s-t-u)-2m_D^2+m_\pi^2)$ \\
\hline
$\eta_c K^0 \rightarrow \pi^0K^0$  & $-{1\over6\sqrt{6}f^2}(-2m_K^2-m_\pi^2)$ \\
$\rightarrow \pi^-K^+$  & $-{1\over6\sqrt{3}f^2}(2m_K^2+m_\pi^2)$ \\
$\rightarrow \eta K^0$  & $-{1\over6\sqrt{2}f^2}(-2m_K^2+m_\pi^2)$ \\
\hline
\end{tabular}

\subsection{C=0, S=0}

\begin{tabular}{|r|c|}
\hline
States & Amplitude \\
\hline
\hline
$D_s^+D_s^- \rightarrow D_s^+D_s^-$ & $-{1\over3f^2}(\psi_3(s+t-2u)+2m_D^2+2m_K^2-2m_\pi^2)$ \\
$\rightarrow D^+D^-$  & $-{1\over6f^2}(\psi_5(t-u)+s-u+2m_D^2+m_K^2-m_\pi^2)$ \\
$\rightarrow D^0\bar D^0$  & $-{1\over6f^2}(\psi_5(t-u)+s-u+2m_D^2+m_K^2-m_\pi^2)$ \\
\hline
$D^+D^-\rightarrow D^+D^-$ & $-{1\over3f^2}(\psi_3(s+t-2u)+2m_D^2)$ \\
$\rightarrow D^0\bar D^0$ & $-{1\over6f^2}(\psi_5(t-u)+s-u+2m_D^2)$ \\
\hline
$D^0\bar D^0\rightarrow D^0\bar D^0$  & $-{1\over3f^2}(\psi_3(s+t-2u)+2m_D^2)$ \\
\hline
$K^+K^- \rightarrow K^+K^-$  & $-{1\over3f^2}(s+t-2u+2m_K^2)$ \\
$\rightarrow K^0\bar K^0$ & $-{1\over6f^2}(s+t-2u+2m_K^2)$ \\
$\rightarrow \pi^+\pi^-$ & $-{1\over6f^2}(s+t-2u+m_K^2+m_\pi^2)$ \\
$\rightarrow \pi^0\pi^0$ & $-{1\over12f^2}(2s-t-u+2m_K^2+2m_\pi^2)$ \\
$\rightarrow \pi^0\eta$ & $-{1\over12\sqrt{3}f^2}(3(2s-t-u)-2m_K^2+2m_\pi^2)$ \\
$\rightarrow \eta\eta$  & $-{1\over12f^2}(3(2s-t-u)+6m_K^2-2m_\pi^2)$ \\
\hline
$K^0\bar K^0\rightarrow K^0\bar K^0$  & $-{1\over3f^2}(s+t-2u+2m_K^2)$ \\
$\rightarrow \pi^+\pi^-$ & $-{1\over6f^2}(s-2t+u+m_K^2+m_\pi^2)$ \\
$\rightarrow \pi^0\pi^0$ & $-{1\over12f^2}(2s-t-u+2m_K^2+2m_\pi^2)$ \\
$\rightarrow \pi^0\eta$ & $-{1\over12\sqrt{3}f^2}(-3(2s-t-u)+2m_K^2-2m_\pi^2)$ \\
$\rightarrow \eta\eta$  & $-{1\over12f^2}(3(2s-t-u)+6m_K^2-2m_\pi^2)$ \\
\hline
$\pi^+\pi^-\rightarrow \pi^+\pi^-$  & $-{1\over3f^2}(s+t-2u+2m_\pi^2)$ \\
$\rightarrow \pi^0\pi^0$ & $-{1\over3f^2}(2s-t-u+m_\pi^2)$ \\
$\rightarrow \pi^0\eta$ & -  \\
$\rightarrow \eta\eta$ & $-{1\over3f^2}m_\pi^2$ \\
\hline
$\pi^0\pi^0\rightarrow \pi^0\pi^0$  & $-{1\over f^2}m_\pi^2$ \\
$\rightarrow \pi^0\eta$  & -  \\
$\rightarrow \eta\eta$ & $-{1\over3f^2}m_\pi^2$ \\
\hline
$\pi^0\eta\rightarrow \pi^0\eta$ & $-{1\over3f^2}m_\pi^2$ \\
$\rightarrow \eta\eta$  & -  \\
\hline
$\eta\eta\rightarrow \eta\eta$  & $-{1\over9f^2}(16m_k^2-7m_\pi^2)$ \\
\hline
$\eta_c\pi^0\rightarrow \eta_c\pi^0$  & $-{1\over6f^2}m_\pi^2$ \\
$\rightarrow \eta_c\eta$  & -  \\
\hline
$\eta_c\eta\rightarrow \eta_c\eta$  & $-{1\over18f^2}(4m_k^2-m_\pi^2)$ \\
\hline
$D_s^+D_s^- \rightarrow K^+K^-$  & $-{1\over6f^2}(t-u+\gamma(s-u)+m_D^2+2m_K^2-m_\pi^2)$ \\
$\rightarrow K^0\bar K^0$ & $-{1\over6f^2}(t-u+\gamma(s-u)+m_D^2+2m_K^2-m_\pi^2)$ \\
$\rightarrow \pi^+\pi^-$ & -  \\
$\rightarrow \pi^0\pi^0$ & - \\
$\rightarrow \pi^0\eta$ & -  \\
$\rightarrow \eta\eta$  & $-{1\over9f^2}(\gamma(2s-t-u)+2m_D^2+6m_K^2-4m_\pi^2)$ \\
\hline
\end{tabular}

\begin{tabular}{|r|c|}
\hline
States & Amplitude  \\
\hline
\hline
$D^+D^- \rightarrow K^+K^-$ & -  \\
$\rightarrow K^0\bar K^0$ & $-{1\over6f^2}(-(t-u)+\gamma(s-t)+m_D^2+m_K^2)$ \\
$\rightarrow \pi^+\pi^-$ & $-{1\over6f^2}(t-u+\gamma(s-u)+m_D^2+m_\pi^2)$ \\
$\rightarrow \pi^0\pi^0$ & $-{1\over12f^2}(\gamma(2s-t-u)+2m_D^2+2m_\pi^2)$ \\
$\rightarrow \pi^0\eta$ & $-{1\over12\sqrt{3}f^2}(-\gamma(2s-t-u)-2m_D^2-2m_\pi^2)$ \\
$\rightarrow \eta\eta$ & $-{1\over36f^2}(\gamma(2s-t-u)+2m_D^2+2m_\pi^2)$ \\
\hline
$D^0\bar D^0 \rightarrow K^+K^-$ & $-{1\over6f^2}(-(t-u)+\gamma(s-t)+m_D^2+m_K^2)$ \\
$\rightarrow K^0\bar K^0$ & - \\
$\rightarrow \pi^+\pi^-$ & $-{1\over6f^2}(-(t-u)+\gamma(s-t)+m_D^2+m_\pi^2)$ \\
$\rightarrow \pi^0\pi^0$ & $-{1\over12f^2}(\gamma(2s-t-u)+2m_D^2+2m_\pi^2)$ \\
$\rightarrow \pi^0\eta$ & $-{1\over12\sqrt{3}f^2}(\gamma(2s-t-u)+2m_D^2+2m_\pi^2)$ \\
$\rightarrow \eta\eta$ & $-{1\over36f^2}(\gamma(2s-t-u)+2m_D^2+2m_\pi^2)$ \\
\hline
$D_s^+D_s^-\rightarrow \eta_c\pi^0$ & -  \\
$\rightarrow \eta_c\eta$  & $-{1\over9\sqrt{2}f^2}(-2\gamma(2s-t-u)+2m_D^2-m_\pi^2)$ \\
\hline
$D^+D^-\rightarrow \eta_c\pi^0$  & $-{1\over6\sqrt{6}f^2}(-2\gamma(2s-t-u)+2m_D^2-m_\pi^2)$ \\
$\rightarrow \eta_c\eta$ & $-{1\over18\sqrt{2}f^2}(2\gamma(2s-t-u)-2m_D^2+m_\pi^2)$ \\
\hline
$D^0\bar D^0\rightarrow \eta_c\pi^0$  & $-{1\over6\sqrt{6}f^2}(2\gamma(2s-t-u)-2m_D^2+m_\pi^2)$ \\
$\rightarrow \eta_c\eta$  & $-{1\over18\sqrt{2}f^2}(2\gamma(2s-t-u)-2m_D^2+m_\pi^2)$ \\
\hline
$\eta_c\pi^0\rightarrow K^+K^-$  & $-{1\over6\sqrt{6}f^2}(2m_K^2+m_\pi^2)$ \\
$\rightarrow K^0\bar K^0$ & $-{1\over6\sqrt{6}f^2}(-2m_K^2-m_\pi^2)$ \\
$\rightarrow \pi^+\pi^-$ & -  \\
$\rightarrow \pi^0\pi^0$ & -  \\
$\rightarrow \pi^0\eta$ & $-{1\over3\sqrt{2}f^2}m_\pi^2$ \\
$\rightarrow \eta\eta$  & -  \\
\hline
$\eta_c\eta\rightarrow K^+K^-$ & $-{1\over6\sqrt{2}f^2}(-2m_K^2+m_\pi^2)$ \\
$\rightarrow K^0\bar K^0$ & $-{1\over6\sqrt{2}f^2}(-2m_K^2+m_\pi^2)$ \\
$\rightarrow \pi^+\pi^-$  & $-{1\over3\sqrt{2}f^2}m_\pi^2$ \\
$\rightarrow \pi^0\pi^0$  & $-{1\over3\sqrt{2}f^2}m_\pi^2$ \\
$\rightarrow \pi^0\eta$ & -  \\
$\rightarrow \eta\eta$ & $-{1\over9\sqrt{2}f^2}(-8m_K^2+5m_\pi^2)$ \\
\hline
\end{tabular}

\section{Isospin and SU(3) basis}

 The following phases are taken for the meson assignments of the 15-plet:

 $|D_s>_0=|D_s^+>$, $|D>_{1\over2}=\left( \begin{array}{c} |D^+> \\ -|D^0> \end{array} \right)$, 
$|K>_{1\over2}=\left( \begin{array}{c} |K^+> \\ |K^0> \end{array} \right)$, 

$|\pi>_{1}=\left( \begin{array}{c} -|\pi^+> \\ |\pi^0> \\ |\pi^-> \end{array} \right)$, $|\eta>_0=|\eta>$, 
$|\eta_c>_0=|\eta_c>$, $|\bar K>_{1\over2}=\left( \begin{array}{c} |\bar K^0> \\ -|K^-> \end{array} \right)$, 

$|\bar D>_{1\over2}=\left( \begin{array}{c} |\bar D^0> \\ |D^-> \end{array} \right)$ and $|\bar{D_s}>_0=|D_s^->$

 In the following we will list for the sectors where a $SU(3)$ decomposition is not trivial, the isospin and $SU(3)$
 states used to transform the amplitudes from a charge basis to isospin and then from isospin into a SU(3) basis. For
 reviews on phase conventions and isoscalar factors of the $SU(3)$ Clebsch-Gordan coefficients one can refer to [\ref{su31}],[\ref{su32}].

\subsection{$\bar 3 \otimes \bar 3$ \hspace{3cm} (\bf{C=2})}

$|D_sD_s>_0=|D_s^+ D_s^+>$

$|DD_s>_{1\over2}=-|D^0 D_s^+>$

$\left( \begin{array}{c} |DD>_0 \\ |DD>_1 \end{array} \right) = 
{1\over\sqrt{2}}\left( \begin{array}{cc} -1&1\\-1&-1\end{array} \right) \left( \begin{array}{c} |D^+D^0> \\
 |D^0D^+> \end{array} \right)$

\begin{figure}[h]
\begin{center}
\includegraphics[scale=0.2,angle=-90]{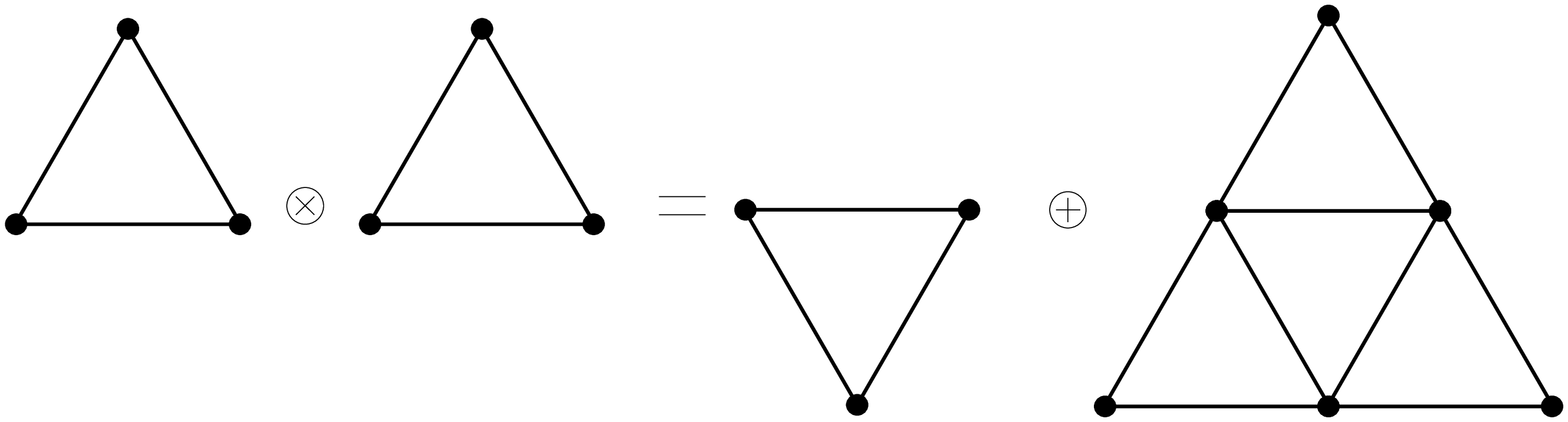} 
\caption{$\bar 3 \otimes \bar 3=3 \oplus \bar 6$.} 
\end{center}
\end{figure}

$|\bar 6,2,0>\footnote{$SU(3)$ states are represented as $|Irrep,S,I>$}=|D_sD_s>_0$

$|\bar 6,1,\oh>=\squd(|DD_s>+|D_sD>)\footnote{from now on the label for the isospin of the states will be omitted for the $SU(3)$ states.}$

$|\bar 6,0,1>=|DD>$

$|3,1,0>=\squd(|DD_s>-|D_sD>)$

$|3,0,\oh>=|DD>$

\subsection{$\bar 3\otimes 8$\hspace{3cm}\bf{C=1}}

$|KD_s>_{1\over2}=|K^0D_s^+>$

$\left( \begin{array}{c} |KD>_0 \\ |KD>_1 \end{array} \right) = 
{1\over\sqrt{2}}\left( \begin{array}{cc} -1&-1\\-1&1\end{array} \right) \left( \begin{array}{c} |K^+D^0> \\
 |K^0D^+> \end{array} \right)$

$|\eta D_s>_0=|\eta D_s^+>$

$|\pi D_s>_1=|\pi^- D_s^+>$

$\left( \begin{array}{c} |\pi D>_{1\over2} \\ |\pi D>_{3\over2} \end{array} \right) = 
\left( \begin{array}{cc} {-1\over\sqrt{3}}&-\sqrt{2\over3}\\-\sqrt{2\over3}&{1\over\sqrt{3}}\end{array}
 \right) \left( \begin{array}{c} |\pi^0D^0> \\ |\pi^-D^+> \end{array} \right)$

$|\eta D>_{1\over2}=-|\eta D^0>$

$|\bar K D_s>_{1\over2}=-|K^-D_s^+>$

$\left( \begin{array}{c} |\bar KD>_0 \\ |\bar KD>_1 \end{array} \right) = 
{1\over\sqrt{2}}\left( \begin{array}{cc} 1&-1\\-1&-1\end{array} \right) \left( \begin{array}{c} |K^-D^+> \\
 |\bar K^0D^0> \end{array} \right)$

\begin{figure}[h]
\begin{center}
\includegraphics[scale=0.2,angle=-90]{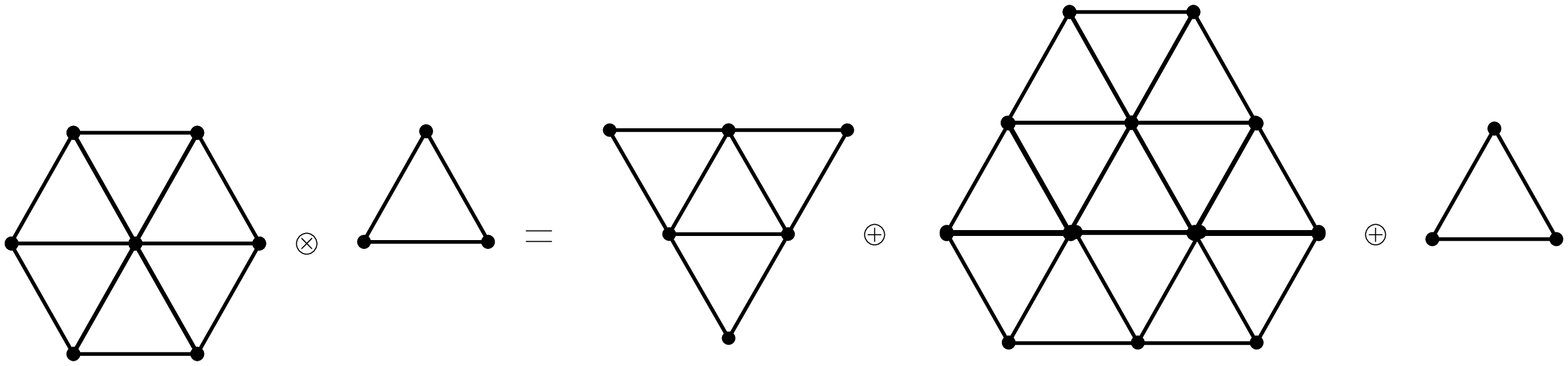} 
\caption{$8\otimes \bar 3 =6 \oplus \bar {15} \oplus \bar 3$.} 
\end{center}
\end{figure}

$|\bar {15},2,\oh>=|KD_s>$

$|\bar {15},1,1>=\squd(|KD>-|\pi D_s>)$

$|\bar{15},1,0>=-{\sqt\over2}|\eta D_s>+\oh|KD>$

$|\bar{15},0,\trh>=|\pi D>$

$|\bar{15},0,\oh>={1\over4}|\pi D>+{3\over4}|\eta D>-\sqrt{3\over8}|\bar KD_s>$

$|\bar{15},-1,1>=|\bar KD>$

$|6,1,1>=\squd(|KD>+|\pi D_s>)$

$|6,0,\oh>=\sqrt{3\over8}|\pi D>-\sqrt{3\over8}|\eta D>-\oh|\bar KD_s>$

$|6,-1,0>=|\bar KD>$

$|\bar3,1,0>=\oh|\eta D_s>+{\sqt\over2}|KD>$

$|\bar3,0,\oh>=-{3\over4}|\pi D>-{1\over4}|\eta D>-\sqrt{3\over8}|\bar KD_s>$

\subsection{$\bar 3 \otimes 3$ \hspace{3cm} \bf{C=0}}

$|D_s\bar D>_\oh=|D_s^+ D^->$

$|D_s \bar{D_s}>_0=|D_s^+D_s^->$

$\left( \begin{array}{c} |D\bar D>_0 \\ |D\bar D>_1 \end{array} \right) = 
{1\over\sqrt{2}}\left( \begin{array}{cc} 1&1\\1&-1\end{array} \right) \left( \begin{array}{c} |D^+D^-> \\
 |D^0\bar D^0> \end{array} \right)$

\begin{figure}[h]
\begin{center}
\includegraphics[scale=0.2,angle=-90]{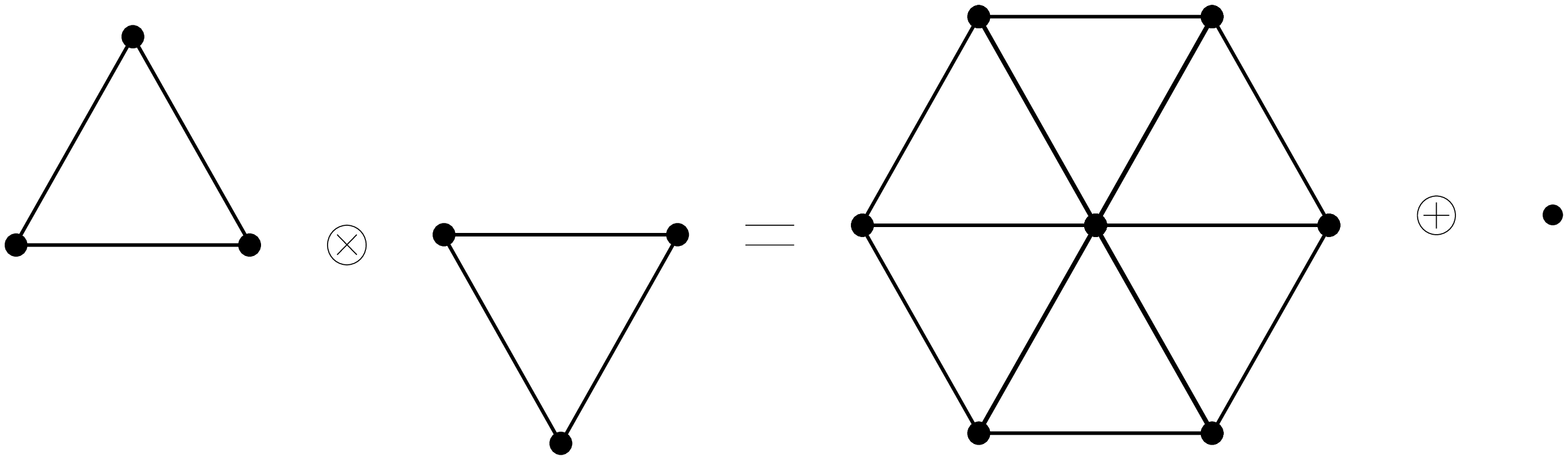} 
\caption{$\bar 3 \otimes 3=8 \oplus 1$.} 
\end{center}
\end{figure}

$|8,1,\oh>=|D_s\bar D>$

$|8,0,1>=|D\bar D>$

$|8,0,0>=\sqdt|D_s\bar D_s>-\squt|D\bar D>$

$|8,-1,\oh>=|\bar D_s D>$

$|1,0,0>=\squt|D_s \bar D_s>+\sqdt|D \bar D>$

\subsection{$8\otimes8$\hspace{3cm}\bf{C=0}}

$\left( \begin{array}{c} |\pi K>_{1\over2} \\ |\pi K>_{3\over2} \end{array} \right) = 
\left( \begin{array}{cc} {1\over\sqrt{3}}&-\sqrt{2\over3}\\ \sqrt{2\over3}&{1\over\sqrt{3}}\end{array}
 \right) \left( \begin{array}{c} |\pi^0K^0> \\ |\pi^-K^+> \end{array} \right)$

$|\eta K>_\oh=|\eta K^0>$

$\left( \begin{array}{c} |K\bar K>_0 \\ |K\bar K>_1 \end{array} \right) = 
{1\over\sqrt{2}}\left( \begin{array}{cc} -1&-1\\-1&1\end{array} \right) \left( \begin{array}{c} |K^+K^-> \\
 |K^0\bar K^0> \end{array} \right)$

$\left( \begin{array}{c} |\pi\pi>_0 \\ |\pi\pi>_1 \\|\pi\pi>_2 \end{array} \right) = 
\left( \begin{array}{ccc} -\squt &-\squt&-\squt \\-\squd&\squd&0 \\ -\squs&-\squs & \sqdt \end{array} \right)
 \left( \begin{array}{c} |\pi^+\pi^-> \\|\pi^-\pi^+> \\ |\pi^0\pi^0> \end{array} \right)$

$|\pi\eta>_1=|\pi^0 \eta>$

\begin{figure}[h]
\begin{center}
\includegraphics[scale=0.2,angle=-90]{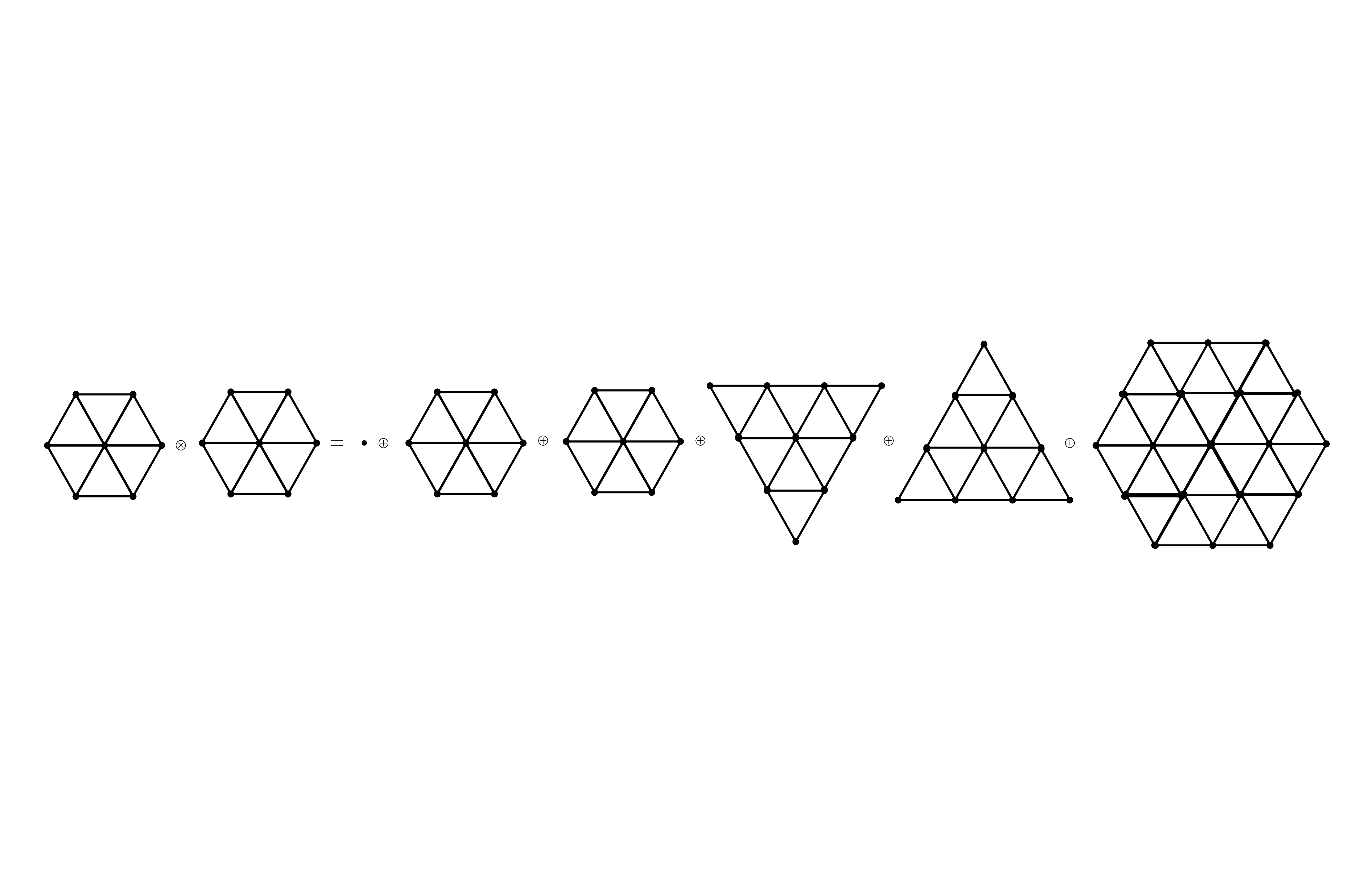} 
\caption{$8\otimes8=1\oplus 8\oplus 8\oplus 10 \oplus\bar{10}\oplus27$} 
\end{center}
\end{figure}

$|27, 2, 1>=|KK>$

$|27, 1 ,\trh>=\squd(|K\pi>+|\pi K>)$

$|27 ,1 ,\oh>={1\over2\sqrt{5}}(|K\pi>+|\pi K>-3|K\eta>-3|\eta K>)$

$|27, 0, 2>=|\pi\pi>$

$|27, 0, 1>={1\over\sqrt{5}}(|K\bar K>+|\bar KK>)+{3\over\sqrt{30}}(|\pi\eta>+|\eta\pi>)$

$|27 ,0 ,0>={3\over2\sqrt{15}}(-|K\bar K>-|\bar KK>)+{1\over2\sqrt{10}}|\pi\pi>-{9\over2\sqrt{30}}|\eta\eta>$

$|\bar{10}, 1, \oh>=\oh(|K\pi>-|\pi K>+|K\eta>-|\eta K>)$

$|\bar{10}, 0, 1>=\squs(|K\bar K>-|\bar KK>-|\pi\pi>)+\oh(|\pi\eta>-|\eta\pi>)$

$|10 ,1, \trh>=\squd(-|K\pi>+|\pi K>)$

$|10, 0, 1>=\squs(-|K\bar K>+|\bar KK>+|\pi\pi>)+\oh(|\pi\eta>-|\eta\pi>)$

$|8_S, 1 ,\oh>={1\over2\sqrt{5}}(-3|K\pi>-3|\pi K>-|K\eta>-|\eta K>)$

$|8_S ,0 ,1>={3\over\sqrt{30}}(-|K\bar K>-|\bar KK>)+{1\over\sqrt{5}}(|\pi\eta>+|\eta\pi>)$

$|8_S, 0 ,0>={1\over\sqrt{10}}(-|K\bar K>-|\bar KK>)+{3\over\sqrt{15}}|\pi\pi>+{1\over\sqrt{5}}|\eta\eta>$

$|8_A, 1, \oh>=\oh(-|K\pi>+|\pi K>+|K\eta>-|\eta K>)$

$|8_A ,0, 1>=\squs(|K\bar K>-|\bar KK>+2|\pi\pi>)$

$|8_A, 0 ,0>=\squd(-|K\bar K>+|\bar KK>)$

$|1 ,0 ,0>=\oh(-|K\bar K>-|\bar KK>)-{3\over2\sqrt{6}}|\pi\pi>+{1\over2\sqd}|\eta\eta>$

\section{Isolating The $J_\psi$ Contribution from $\lc_3$}

 The $\lc_3$ Lagrangian in eq. (\ref{lag3}) has two terms, the first one contains just hadronic currents where the initial and final
 state have the same electric charge and, therefore, exchange neutral vector mesons only. The other term has contributions from both charged and neutral vector mesons; from this second term first one should isolate the contribution from neutral vector mesons. Let us add and subtract the appropriate term, 

\mbox{ }

$Tr\left( J_{3\bar3\mu}J_{3\bar3}^\mu \right) \rightarrow Tr\left( J_{3\bar3\mu}J_{3\bar3}^\mu \right)-\Big( J_{D_s\bar D_s\mu}J_{D_s\bar D_s}^\mu+J_{D^+D^-\mu}J_{D^+D^-}^\mu+J_{D^0\bar D^0\mu}J_{D^0\bar D^0}^\mu\Big)+$

\begin{equation}
 \hspace{2.3cm} +\Big( J_{D_s\bar D_s\mu}J_{D_s\bar D_s}^\mu+J_{D^+D^-\mu}J_{D^+D^-}^\mu+J_{D^0\bar D^0\mu}J_{D^0\bar D^0}^\mu\Big) \label{firstterml3}
\end{equation}

\mbox{ }

 such that now the sum of the first two terms in eq. (\ref{firstterml3}) has no contribution from heavy vector meson which is now in the third term alone.

 The second term of Lagrangian $\lc_3$ in eq. (\ref{lag3}) will then be expanded in order to identify terms where equal hadronic currents are connected and terms where different ones are connected:

\mbox{ }

$J_{\bar3 3\mu}J_{\bar3 3}^\mu=2\Big(J_{D_s\bar D_s\mu}(J_{D^+ D^-}^\mu+J_{D^0 \bar D^0}^\mu)+J_{D^+D^-\mu}J_{D^0\bar D^0}^\mu\Big)+$

$\hspace{1.6cm}  J_{D_s\bar D_s\mu}J_{D_s\bar D_s}^\mu+J_{D^+D^-\mu}J_{D^+D^-}^\mu+J_{D^0\bar D^0\mu}J_{D^0\bar D^0}^\mu$

\mbox{ }

 Now terms with the product of equal neutral hadronic currents are to be multiplied by the correction $\psi_3$ and terms 
 connecting different ones by $\psi_5$, given in eqs. (\ref{psi3fac}) and (\ref{psi5fac}). As a result:

\begin{eqnarray}
\lc_3&=&Tr\left( J_{3\bar3\mu}J_{3\bar3}^\mu \right)-\Big( J_{D_s\bar D_s\mu}J_{D_s\bar D_s}^\mu+J_{D^+D^-\mu}J_{D^+D^-}^\mu+J_{D^0\bar D^0\mu}J_{D^0\bar D^0}^\mu\Big)+ \nonumber \\
& & 2\psi_5\Big(J_{D_s\bar D_s\mu}(J_{D^+ D^-}^\mu+J_{D^0 \bar D^0}^\mu)+J_{D^+D^-\mu}J_{D^0\bar D^0}^\mu\Big)+2\psi_3\Big(
 J_{D_s\bar D_s\mu}J_{D_s\bar D_s}^\mu+\nonumber \\
& & J_{D^+D^-\mu}J_{D^+D^-}^\mu+J_{D^0\bar D^0\mu}J_{D^0\bar D^0}^\mu\Big)
\end{eqnarray}

 One can work it out:

\begin{eqnarray}
\lc_3&=& Tr\left( J_{3\bar3\mu}J_{3\bar3}^\mu \right)+ 2\psi_5\Big(J_{D_s\bar D_s\mu}(J_{D^+ D^-}^\mu+J_{D^0 \bar D^0}^\mu)+J_{D^+D^-\mu}J_{D^0\bar D^0}^\mu\Big)+\nonumber \\
& & \underbrace{(2\psi_3-1)}_{\psi_5} \Big( J_{D_s\bar D_s\mu}J_{D_s\bar D_s}^\mu+ J_{D^+D^-\mu}J_{D^+D^-}^\mu+J_{D^0\bar D^0\mu}J_{D^0\bar D^0}^\mu\Big) = \nonumber \\
& & Tr\left( J_{3\bar3\mu}J_{3\bar3}^\mu \right)+ \psi_5\Big(J_{D_s\bar D_s\mu}J_{D_s\bar D_s}^\mu+J_{D^+D^-\mu}J_{D^+D^-}^\mu+J_{D^0\bar D^0\mu}J_{D^0\bar D^0}^\mu+ \nonumber \\
& & 2J_{D_s\bar D_s\mu}(J_{D^+ D^-}^\mu+J_{D^0 \bar D^0}^\mu)+2J_{D^+D^-\mu}J_{D^0\bar D^0}^\mu\Big) = Tr\left( J_{3\bar3\mu}J_{3\bar3}^\mu \right)+\psi_5 J_{\bar3 3\mu}J_{\bar3 3}^\mu \nonumber
\end{eqnarray}

 And this is the simple form we write down in eq. (\ref{lagfull}). Yet, in the amplitudes we use the $\psi_3$ and $\psi_5$ factors.

\end{document}